\documentclass[english,superscriptaddress, aps, showkeys, prd, floatfix, 11pt, nofootinbib,longbibliography]{revtex4-2}

\usepackage{graphicx}  
\usepackage{amsmath}  
\usepackage{amssymb}  
\usepackage{caption}  
\usepackage{subcaption}
\usepackage[normalem]{ulem}  
\usepackage{orcidlink}  

\DeclareMathOperator{\arcsech}{arcsech}  
\DeclareMathOperator{\sech}{sech}  

\count\footins = 1000 

\graphicspath{ {./Figures/} }

\usepackage{xcolor}  

\newcommand{\be}{\begin{equation}}  
\newcommand{\ee}{\end{equation}}  

\newcommand{\pphi}{p_\phi}  
\newcommand{\ppsi}{p_\psi}  
\newcommand{\pphiB}{p_{\phi B}}  
\newcommand{\ppsiB}{p_{\psi B}}  
\newcommand{\rhoc}{\rho_c}  
\newcommand{\Qphi}{\mathcal{Q}^{\phi}}  
\newcommand{\Qpsi}{\mathcal{Q}^{\psi}}  
\newcommand{\QF}{\mathcal{Q}^{F}}  
\newcommand{\Rcurv}{\mathcal{R}}  
\newcommand{\VF}{\mathcal{V}^F}   
\newcommand{\PSphi}{\mathcal{P}_\phi} 
\newcommand{\PSpsi}{\mathcal{P}_\psi} 
\newcommand{\PSR}{\mathcal{P}_\Rcurv} 
\newcommand{\PSS}{\mathcal{P}_\mathcal{S}} 
\newcommand{\PST}{\mathcal{P}_t} 

\begin{document}

\title{Quasi-dust ekpyrotic scenario in Loop Quantum Cosmology}

\author{Emmanuel Frion
\orcidlink{0000-0003-1280-0315}}
\email{efrion@uwo.ca}
\affiliation{Dept.\,of Physics \& Astronomy, Western University, N6A\,3K7 London ON, Canada}

\author{Mateo Pascual
\orcidlink{0000-0001-7706-3075}}
\email{mpascua@uwo.ca}
\affiliation{Dept.\,of Physics \& Astronomy, Western University, N6A\,3K7 London ON, Canada}

\author{Francesca Vidotto
\orcidlink{0000-0002-9883-0808}}
\email{fvidotto@uwo.ca}
\affiliation{Dept.\,of Physics \& Astronomy, Western University, N6A\,3K7 London ON, Canada}
\affiliation{Dept.\,of Philosophy and Rotman Institute, Western University, N6A\,3K7 London ON, Canada}
\affiliation{Instituto de Estructura de la Materia, IEM-CSIC,
28006 Madrid, Spain}


\begin{abstract}
In the framework of Loop Quantum Cosmology, we study a cosmological bouncing model with two fields that reproduce the desired features of the primordial power spectra.
The model combines the matter-bounce mechanism, that generates scale-invariant perturbations, with ekpyrotic contraction, that suppresses anisotropies leading to the bounce.
The bounce that replaces the classical initial singularity is achieved thanks to the loop quantisation.
The matter-bounce is enacted by a \textit{quasi-dust} scalar field, with a slightly-negative equation of state that accounts for a small positive cosmological constant, that generates a red-tilt in the perturbations' power spectra.
A second field, endowed with an ekpyrotic potential, is introduced to tame the growth of anisotropies throughout the bouncing phase.
The equations of motion of the scalar perturbations are non-trivially coupled, leading to rich phenomenology that cannot be inferred simply from their single-field counterpart.
We study the evolution of scalar and tensor perturbations and compare the results to current observations, showing the viability of this model as a base for further investigations.
\end{abstract}

\maketitle

\section{Introduction}

The Standard Model of cosmology typically includes a phase of accelerated expansion -- cosmological inflation 
\cite{Starobinsky:1980te, Sato:1980yn, Guth:1980zm, Linde:1981mu, Albrecht:1982wi, Linde:1983gd, Mukhanov:1981xt, Mukhanov:1982nu, Starobinsky:1982ee, Guth:1982ec, Hawking:1982cz, Bardeen:1983qw} %
-- to account for key features of the cosmic microwave background (CMB)%
\cite{WMAP:2012nax, Planck:2018nkj, Planck:2018vyg, Planck:2018jri}. The model breaks down at very small scales %
\cite{Brandenberger:2012aj}, motivating the search for alternative models that address the initial conditions of the Standard Model and resolve the problem of the initial singularity.

These scales are expected to be the domain of a quantum theory of gravity. Among the different proposals, 
Loop Quantum Gravity (LQG) succeeds in removing the classical curvature singularities by combining the presence of a minimum non-vanishing eigenvalue for geometrical quantities and the boundedness of curvature \cite{Rovelli:1994ge, Rovelli:2013osa, Li:2023dwy, ElizagaNavascues:2023vll}. Loop Quantum Cosmology (LQC) studies the cosmological phenomenology that can be derived from the Hamiltonian dynamics of this theory, leading to the prediction of a Big Bounce replacing the traditional Big Bang singularity, happening when the energy density of the universe reaches the Planck density $\rho_{Pl}$  
\cite{Ashtekar:2006rx}. We refer the reader to the  recent reviews for a full description of this framework (see for instance \cite{Agullo:2016tjh, ElizagaNavascues:2020uyf, Barca:2021qdn, Agullo:2023rqq}).

While most of the LQC literature focuses on scenarios where an inflationary phase follows the quantum bounce %
\cite{Ashtekar:2009mm, Bonga:2015kaa, Bhardwaj:2018omt}, the non-singular nature of the bounce provides a natural arena for the development of alternative early-universe models. 
The appeal of inflation resides in its ability to generate scale-invariant primordial perturbations, but it is known that these can also arise through different mechanisms.
Among these, the ``matter bounce'' just requires a phase of pressure-less matter (dust) domination during the contracting phase of a bounce to produce 
scale-invariant power spectra %
\cite{Wands:1998yp, Brandenberger:2012zb}.

In \cite{Wilson-Ewing:2012lmx}, Wilson-Ewing developed a bouncing scenario from LQC using a perfect fluid with a constant equation of state. In the case of a dust field, with vanishing equation of state, this LQC matter-bounce scenario predicts scale-invariant power spectra of scalar and tensor perturbations. It was observed in \cite{Wilson-Ewing:2012lmx} that a slightly-negative equation of state would reproduce the red-tilted scalar power spectrum observed in the CMB (see also \cite{Peter:2006hx, Peter:2008qz, Miranda:2019ara}).
However, the amplitude of scalar perturbations suggests a critical energy density significantly lower than the Planck energy density, approximately $\rhoc \approx 10^{-9} \rho_{Pl}$. Besides this intriguing feature, 
the matter bounce scenario alone also suffers from the Belinski-Khalatnikov-Lifshitz (BKL) instability during the contraction phase \cite{Belinsky:1970ew}.

Li, Saini and Singh analysed a matter-bounce scenario \cite{Li:2020pww} induced by a constant presureless dust component, to which they introduced an additional stiff field, with an \emph{ekpyrotic} potential \cite{Khoury:2001bz,Khoury:2001wf}, to mitigate the BKL instability. Furthermore, such potential provides a natural homogenisation mechanism \cite{Brown:2024xta}. However, the LQC matter-ekpyrotic bouncing scenario predicts an exactly scale-invariant power spectrum, which does not align with observations, necessitating further refinement. 

Here we develop a two-field LQC matter-ekpyrotic bounce that accounts for both the spectral index and the amplitude of scalar and tensor perturbations. The model uses the same ekpyrotic potential as in \cite{Li:2020pww}, varying the associated parameters. We design the matter field to act as quasi-dust when dominant in the far past and future of the bounce. Consequently, we dub it as the \textit{quasi-dust} field throughout the paper, even though its characteristics evolve over time. We solve numerically the equations for the evolution of the background for the perturbations.

The manuscript is organised as follows. In section \ref{sec:Background_Dynamics}, we review the background cosmological setting, introducing the formalism for the dynamics of Loop Quantum Cosmology in presence of the quasi-dust and the ekpyrotic fields, and solving it numerically. We discuss how the different choices of parameters lead to different scenarios. 

In section \ref{sec:scalar_perturbations} we study the dynamics of scalar perturbations and their resulting power spectra. 
We motivate the choice of initial perturbations during the quasi-dust-dominated phase of contraction. 
We then thoroughly study the evolution of the power spectra of curvature and entropy perturbations for a wide range of modes, as well as the amplitude of two modes that represent those that are larger or smaller than the Hubble radius initially for the time-span explored. We close this section by discussing the effects of different parameter choices on the scalar power spectra.

In section \ref{sec:Dynamics_Tensor_perturbations} we carry out a similar procedure for tensor perturbations as mentioned for the scalar sector. We close with a conclusion and outlook into further investigations.

We take a closer look at a peculiar behaviour of scalar perturbations in appendix~\ref{app:real-imaginary-components}, and include the plots for the evolution of perturbations at the pivot scale in appendix~\ref{app:kPivot}.

We use natural units with $\hbar=c=G=1$. In the formulas, we keep the Newton’s constant $G$ explicit, while in the numerical solutions $G$ is also set to unity. We also invite the reader to pay special attention to the \textit{signed-logarithmic} scale used for time in horizontal axes as well as in some vertical axes of several plots throughout the paper.

\section{Background dynamics of the quasi-dust-ekpyrotic bounce in LQC}
\label{sec:Background_Dynamics}

For the matter-ekpyrotic scenario to be a plausible account for our current cosmological observations, one needs to consider a cosmological setting that smoothly patches a contracting universe with an expanding one. In this setting, scale-invariant perturbations generated in a contracting era can propagate smoothly through a bounce into an expanding era preserving their scale-invariant profile. In LQC, the underlying quantum geometry generically results in a bounce taking place when the energy density reaches the Planck scale, providing a natural setting to study the matter-ekpyrotic scenario.

In this section, we study the effective background dynamics of the spatially flat, homogeneous and isotropic loop quantum cosmology with two scalar fields that have specific characteristics.
One scalar field will be tailored so that, when it is the dominant energy component of the universe, it mimics the behaviour of pressureless dust field combined with a small, yet non-negligible, dark energy component. 
The other scalar field will be an Ekpyrotic scalar field, which is necessary to ensure anisotropies are adequately tamed leading to the bounce phase, avoiding the BKL instability. We consider both scalar fields to be minimally coupled to gravity. We will use the Hamilton's equations for the background dynamics to numerically solve for background solutions, setting initial conditions for the background at the bounce time.

\subsection{Effective dynamics of LQC with quasi-dust and ekpyrotic scalar fields}

LQC is a finite, symmetry-reduced sector of LQG that emerges from the canonical quantisation of symmetry-reduced cosmological models using LQG techniques \cite{Ashtekar:2011ni, Bojowald:2008zzb}. 
In this framework, the classical Hamiltonian constraint is reformulated in terms of the Ashtekar-Barbero connection and its conjugate triad. 
For a spatially flat FLRW universe, these variables can in turn be reduced to the canonical pair $c$ and $p$ by virtue of the underlying symmetry.
Once the Hamiltonian formulation of classical spacetimes has been suitably reexpressed, it can be quantised using the so-called $\bar{\mu}$ scheme. 
This results in a non-singular, quantum difference equation with equal steps in volume \cite{Ashtekar:2006wn}. 
The discrete quantum evolution of sharply peaked states in LQC has been shown to be faithfully captured by the effective dynamics of LQC \cite{Agullo:2016tjh}. 
Consequently, it is more practical to employ these effective dynamics when studying the phenomenology arising from the quantum nature of spacetime and from linear perturbations of the background.
The effective dynamics is prescribed by an effective Hamiltonian constraint defined on a phase space that includes both gravitational and matter degrees of freedom. 
Owing to the homogeneity and isotropy of a spatially flat FLRW universe, the gravitational sector can be recast in terms of the canonical pair $(b, v)$, where $v = |p|^{3/2}$ and $b = c\,|p|^{-1/2}$, with the Poisson bracket $\{b,v\} = 4 \pi G \gamma$, and $\gamma$ denoting the Barbero-Immirzi parameter.
In our numerical work, this parameter is fixed at $\gamma = 0.2375$, a value motivated by black hole thermodynamics and commonly adopted in the LQC literature \cite{Meissner:2004ju,Agullo:2016tjh}. 
In this work we will be investigating a model in which the matter sector is comprised of an \textit{ekpyrotic} scalar field $\phi$ and its conjugate momentum $\pphi$, with Poisson bracket $\left\{ \phi , \pphi \right\} = 1$, and a \textit{quasi-dust} scalar field $\psi$ and its conjugate momentum $\ppsi$, with Poisson bracket $\left\{ \psi , \ppsi \right\} = 1$.

In light of the cosmological setting outlined above, the effective Hamiltonian that describes the dynamics of this loop-quantised, spatially-flat Friedmann-Lemaître-Ro\-bert\-son-Walker (FLRW) model is given by \cite{Ashtekar:2006wn}
\be \label{eq:effHam}
\mathcal{H} = - \rhoc v \sin^2(\lambda b) + \mathcal{H}_m \;,
\ee
where $\lambda = \sqrt{\Delta}$ with $\Delta = 4\sqrt{3} \pi \gamma l_{Pl}^2$ representing the minimum area eigenvalue in LQG, where the Planck length $l_{Pl}=1$ in natural units,  and $\rho_c = 3/(8\pi G\gamma^2\lambda^2) = 0.41 \rho_{Pl}$ is the critical energy density in LQC at which the bounce takes place, for which the Planck energy density $\rho_{Pl}=1$ in natural units. The effective Hamiltonian is constrained to vanish identically.

The matter Hamiltonian $\mathcal{H}_m$ consists of the ekpyrotic scalar field $\phi$ with potential $U(\phi)$, the quasi-dust scalar field $\psi$ with potential $V(\psi)$, and their respective conjugate momenta $\pphi$ and $\ppsi$.
It takes the form
\be
\mathcal{H}_m = \frac{\pphi^2}{2v} + v U(\phi) + \frac{\ppsi^2}{2v} + v V(\psi) \quad .
\ee

We take the quasi-dust potential to be
\be \label{eq:Quasi-dustPotential}
V(\psi) = V_0 \sech^2 (A \psi) \;,
\ee
where $V_0 = \frac{\rhoc (1-\epsilon)}{2}$ and $A = \sqrt{6 \pi G (1-\epsilon)}$. This choice of potential ensures that when the quasi-dust scalar field dominates, the cosmology is that of a perfect fluid with a constant equation of state given by $P=-\epsilon \rho$, where $0<\epsilon \ll 1$ \cite{Wilson-Ewing:2012lmx, Mielczarek:2008qw}. 

\begin{figure}[tb!]
    \centering
    \makebox[\linewidth]{%
        \includegraphics[width=0.52\textwidth]{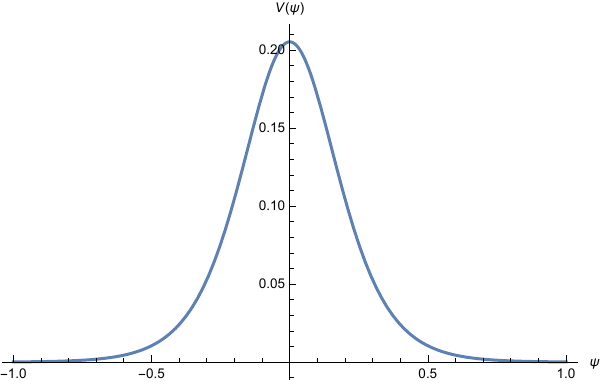}
        \includegraphics[width=0.52\textwidth]{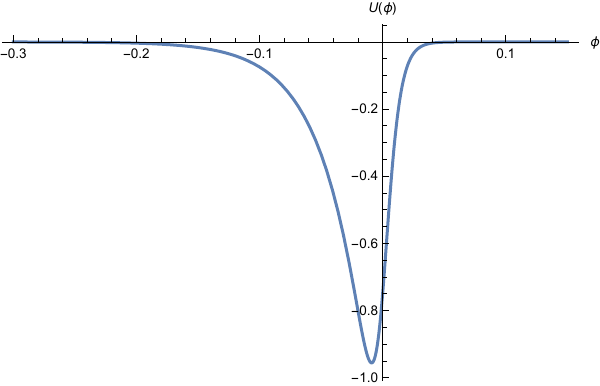}
    }
    \caption{(Left) Plot of the quasi-dust potential $V(\psi)$ for the choice of parameter $\epsilon = 0.0029$ given in equation \eqref{eq:epsilon}.
    This parameter controls both the width and the height of the potential.
    (Right) Plot of the ekpyrotic potential $U(\phi)$ for the choice of parameters $U_0 = 0.75$, $\alpha = 30$ and $\beta = 5$ given in equation \eqref{eq:EkpyroticParameters}. 
    The parameters $U_0$, $\alpha$ and $\beta$ control the depth, width and asymmetry of the ekpyrotic potential respectively.
    }
    \label{fig:quasi-dust-ekpyrotic-potentials}
\end{figure}

This choice of equation of state parameter mimics a cosmological setting that is majorly dominated by pressureless dust matter with a vanishing equation of state, yet also has a small but non-negligible contribution from dark energy, characterised by an equation of state of $-1$. The resulting effective equation of state of this setting would be a negative value that is very close to zero, namely $-\epsilon$, and this is effectively captured by the dynamics of the quasi-dust field when it dominates the dynamics. This near-zero-yet-negative effective equation of state during the contracting phase of the bouncing cosmology is what gives rise to a red-tilt in the otherwise scale-invariant primordial power spectrum of adiabatic scalar perturbations, as will be shown in section \ref{sec:scalar_perturbations}.
The quasi-dust potential given in \eqref{eq:Quasi-dustPotential} is positive definite, features a single maximum point at 
\be
\psi_{max}=0 \, , \qquad \text{ for which }\qquad V_{max}=V(\psi_{max})=\frac{\rhoc (1-\epsilon)}{2}
\ee
and quickly goes to 0 as $\psi \rightarrow \pm \infty$, as can be seen in the left panel in Figure \ref{fig:quasi-dust-ekpyrotic-potentials}.


It is worth pointing out that the factor of $\rhoc$ in \eqref{eq:Quasi-dustPotential} could actually be replaced by any value between zero and $\rhoc$, and the field would still behave as quasi-dust when dominant 
\footnote{It is really only limited to $\rhoc$ in the LQC scenario with only one perfect-fluid scalar field. In our model, with two scalar fields, in principle we could even chose a value higher than $\rhoc$ as long as we ensure that the quasi-dust field is subdominant near and through the bounce at a potential value significantly below $\rho_c$. This is because when this field is subdominant it does not behave like a perfect-fluid anyway, as it freezes due to the dynamics of the dominant ekpyrotic field. Nevertheless, it seems physically unreasonable to model a potential where this factor is larger than the maximum energy density allowed by the underlying geometry, hence we will not entertain this possibility in this work.}
. We choose this factor to be $\rhoc$ because near the time of equality in domination of fields, a larger gradient of the potential results in a quicker transition to perfect fluid behaviour, and the gradient of the potential is proportional to this factor. Hence, this choice results in the fastest recovery of perfect fluid behaviour. 

Furthermore, we take the ekpyrotic potential, as suggested in \cite{Cai:2012va}, to be
\be \label{eq:EkpyroticPotential}
U(\phi) = \frac{-2 \, U_0}{e^{-\alpha \, \phi} + e^{\,\beta \, \alpha \, \phi}} \;,
\ee
where $U_0$, $\alpha$,%
\footnote{The original formulation of this potential in \cite{Cai:2012va} uses $p$ to parametrise the width of the ekpyrotic potential. The different parameters are simply related by $\alpha := \sqrt{\frac{16\pi G}{p}}$. Their motivation originated in the fact that in a classical cosmological setting with a scalar field with a potential given by $U(\phi) \propto \exp \left[ \sqrt{24 \pi G (1+w)} \phi \right]$, the cosmology is that of a perfect-fluid with a constant equation of state $w$, and the scale factor evolves as a power law $a(t) \propto t^p$, where $p = \frac{2}{3(1+w)}$. However we prefer to parametrise the width with $\alpha$ to keep the expression cleaner.
}
and $\beta$ are the three ekpyrotic parameters associated to the depth, width, and asymmetry of the potential respectively, and are all positive. 
The ekpyrotic potential is negative definite, features a single minimum at 
\be
\phi_{min} = -\frac{\ln(\beta)}{\alpha (1+\beta)} \text{,\quad for which} \quad
U_{min} = U(\phi_{min}) =  -\frac{2\, U_0}{1+\beta}\,\beta^{\frac{\beta}{1+\beta} \;} \, ,
\ee
and quickly goes to 0 as $\phi \rightarrow \pm \infty$, as can be seen in the right panel of Figure \ref{fig:quasi-dust-ekpyrotic-potentials}.
Due to the negative exponential potential resembling a well, the ekpyrotic field will exhibit an \emph{ultra-stiff} equation of state parameter (i.e. $w>1$) as the field swiftly traverses the potential well. We will impose this to happen close to the bounce by setting appropriate initial conditions.

From the effective Hamiltonian constraint \eqref{eq:effHam}, we derive the Hamilton equations of motion and find
\begin{align}
    \dot{b} &= -\frac{3}{2\gamma\lambda^2} \sin^2(\lambda b) - 4\pi G \gamma P \label{eq:bdot} \;, \\
    \dot{v} &= \frac{3}{2\gamma\lambda} \sin(2\lambda b) v \label{eq:vdot} \;, \\
    \dot{\phi} &= \frac{\pphi}{v} \label{eq:phidot} \;, \\
    \dot{\pphi} &= -v U_{,\phi} \label{eq:pphidot} \;, \\
    \dot{\psi} &= \frac{\ppsi}{v} \label{eq:psidot}\\
    \dot{\ppsi} &= -v V_{,\psi}  \;, \label{eq:ppsidot}
\end{align}
where a dot above a variable denotes differentiation with respect to proper time $t$, and a comma followed by a field in the subscript of a potential denotes the derivative of that potential with respect to the field. Furthermore, $P$ represents the isotropic pressure defined as $P=-\frac{\partial \mathcal{H}_m}{\partial v}$. This system of equations is comprised of six intricately coupled differential equations describing the evolution of the homogeneous and isotropic background universe. We will solve the background system numerically in the following section.

Using equations \eqref{eq:vdot}, \eqref{eq:phidot} and \eqref{eq:psidot} , as well as the vanishing Hamiltonian constraint \eqref{eq:effHam}, we can show that the background satisfies the modified Friedmann equation in LQC
\be \label{eq:LQCFriedmann}
    H^2 = \frac{8 \pi G}{3} \rho \left( 1 - \frac{\rho}{\rhoc} \right) \;.
\ee
where $H =\frac{\dot{a}}{a}$ denotes the Hubble parameter, and $a=v^{1/3}$ is the scale factor. As $\rho \rightarrow \rhoc$ from below, the Hubble parameter vanishes and reverses sign, indicating a cosmological bounce. This modified Fried\-mann equation is generic in LQC, and the bounce occurs independently of the matter sector of the model
\footnote{In the literature of ekpyrotic cosmology it is common to find models in which the cosmological bounce appearing in the model is generated by an ekpyrotic potential like the one we employ in this work, along with a ghost condensate mechanism \cite{Lin:2010pf}. However, we want to remark that in this model the bounce takes place due to the quantum gravity corrections captured in the LQC modified Friedmann equation independently of the matter sector, and not due to an exotic mechanism related to the ekpyrotic potential.}.

The continuity equation for the matter fields, which furthermore applies to each individual component, is derived from equations \eqref{eq:phidot} – \eqref{eq:ppsidot} to be the same as in the classical setting and is given by
\be \label{eq:Rho-evolution}
    \dot{\rho} + 3H(\rho+P) = 0 \;.
\ee
Here, the total energy density $\rho$ and total pressure $P$ are defined as the sum of the corresponding quantities for each of the fields as
\begin{gather} \label{eq:total-energy-density-pressure}
    \rho = \rho_\phi + \rho_\psi = \frac{1}{2} \dot{\phi}^2 + U(\phi) + \frac{1}{2} \dot{\psi}^2 + V(\psi) \;, \\
    P = P_\phi + P_\psi = \frac{1}{2} \dot{\phi}^2 - U(\phi) + \frac{1}{2} \dot{\psi}^2 - V(\psi)  \;. \label{eqPressure}
\end{gather}
Expressed in terms of the fields, the continuity equation for each component can be furthermore recast as the Klein-Gordon equations for each field
\be \label{eq:K-Geq}
    \ddot{\phi} + 3 H \dot{\phi} + \frac{\partial U(\phi)}{\partial \phi} = 0 
    \quad \text{and} \quad
    \ddot{\psi} + 3 H \dot{\psi} + \frac{\partial V(\psi)}{\partial \psi} = 0 \, .
\ee
Lastly, the total equation of state parameter and that of each of the fields are denoted by
\be \label{eq:EoS}
w_T = \frac{P}{\rho}
\,\, \text{,} \quad
w_\phi = \frac{P_\phi}{\rho_\phi}
\quad \text{and} \quad
w_\psi = \frac{P_\psi}{\rho_\psi}
\ee

At this point we remark that whenever either field dominates the dynamics of the background cosmology, the evolution of the geometry will simply resemble that of a scenario with only this field present. Importantly, the effective equation of state of the universe at a time in which one of the fields is dominant will be approximately the same as the equation of state exhibited by that field at that time.
In order to generate a red-tilted scalar power spectrum via the usual matter-bounce mechanism, the quasi-dust field must be unequivocally dominant during a sufficiently long phase of contraction. We will ensure this is realised by choosing initial conditions and parameter values appropriately.
In the next section we will proceed to numerically solve for solutions of the background dynamics, and explore the effect of different choices of parameter values and initial conditions set at the bounce.

\subsection{Numerical evolution of background dynamics}

The six Hamilton equations \eqref{eq:bdot}-\eqref{eq:ppsidot} form an intricately coupled system of differential equations. Given its complexity, we solve this system numerically with \textit{Mathematica}
to find solutions for the background dynamics spanning a specified time range%
\footnote{Our Mathematica notebook can be found at \href{https://github.com/EFrion/Quasi-Dust-Ekpyrotic-Bouncing-Model}{https://github.com/EFrion/Quasi-Dust-Ekpyrotic-Bouncing-Model}}%
. All the solutions exhibit a non-singular bounce, which is a generic feature of LQC, and for various choices of parameters the solutions may even exhibit multiple non-singular bounces in the Planck regime. We find this to be a consequence of the complicated background dynamics arising from the interplay between multiple scalar fields. Ultimately, our goal is to study how perturbations propagate from a macroscopic contracting branch to a macroscopic expanding branch through only one non-singular bounce induced by quantum gravity. Hence, in our numerical study we restrict our consideration to parameter values that result in a background solution depicting a universe that starts off in a macroscopic, contracting phase dominated by quasi-dust, it then becomes dominated by an ekpyrotic field, undergoes a bounce in a Planckian regime during ekpyrotic domination, and continues into a macroscopic expanding universe within the time span we explore.

\subsubsection{Initial conditions and parameter choices}

We start by considering the contracting phase dominated by quasi-dust in the far past. At this stage the quasi-dust field dominates and behaves like a perfect fluid with a slightly-negative equation of state due to our choice of quasi-dust potential. The total pressure is dominated by the quasi-dust pressure and takes approximately the same negative value. It is therefore not straightforward to deduce whether $b$ monotonically decreases or not just by inspection of equation \eqref{eq:bdot}. Nevertheless, by restricting our focus to the background solutions we portrayed earlier in this section, we find that $b$ does indeed decrease monotonically throughout the explored time span. Consequently, from equation \eqref{eq:vdot} we are able to deduce that the number of bounces taking place throughout the explored time span is given by $[b(t_i)-b(t_f)]\lambda/\pi$, where $t_i$ and $t_f$ denote the initial and final times of the numerical evolution respectively. We perform the numerical evolution over the cosmic time span $t \in [-10^{12},10^{12}]$ so as to ensure that the setting starts off  with a large contracting universe dominated by the quasi-dust field behaving as a perfect fluid with equation of state $w_\psi \approx -\epsilon$.

The initial conditions for the numerical computation are set at the bounce, which we set at $t = t_B = 0$ without loss of generality. 
Hereon, a subscript ‘$B$’ on a variable indicates it is being evaluated at the bounce. In order to solve the Hamilton equations numerically we must specify the initial values of $v_B$, $b_B$, $\phi_B$, $\pphiB$, $\psi_B$ and $\ppsiB$.

We set $b_B = \pi/(2\lambda)$ so that the Hubble parameter vanishes at the bounce, as can be deduced from equation \eqref{eq:vdot}. This value recovers the correct classical limit as the energy density decreases in the expanding branch to the future of the bounce. Without loss of generality, we set the volume at the bounce to $v_B = 1$ since the Hamilton equations remain invariant with respect to a rescaling of the volume and a reciprocal rescaling of the fields' momenta. 

Consequently, the parameter space of initial conditions at the bounce is comprised of three free parameters after recalling that the Hamiltonian constraint \eqref{eq:effHam} is required to vanish. These three free parameters are related to the quasi-dust and ekpyrotic fields. As proposed in \cite{Li:2020pww}, we choose the value of the ekpyrotic field at the bounce to be near the minimum of its potential, setting $\phi_B = 0$. In fact, if $\phi_B$ were not chosen near the bottom of its potential, the field would traverse the well at a time different from $t_B$, and would induce a secondary bounce. However, as noted earlier, solutions featuring multiple bounces are outside the scope of this study.

A distinctive feature of a model with two minimally coupled scalar fields, such as the one presented here, is that either field will evolve in a different fashion depending on whether it is the dominant energy component or not. 
When one field dominates, both this field and the geometry evolve approximately as if this were the only field present. In contrast, the evolution of the subdominant field is severely affected by the background evolution dictated by the dominant field, specifically due to the influence of the Hubble term ($3H\dot{\phi}$ or $3H\dot{\psi}$) in the subdominant field's Klein-Gordon equation \eqref{eq:K-Geq}. This has an important implication for the choice of initial conditions for the quasi-dust field and its momentum.

During the quasi-dust-dominated phase of contraction, the evolution of the universe is approximately that of a scenario with only this field present. Requiring that the quasi-dust field behaves like a perfect fluid with constant equation of state at the beginning of the time span explored will imply that the field starts off climbing up its potential.
Recalling the equation of motion for $\psi$ in \eqref{eq:K-Geq}, the Hubble term $3H\dot{\psi}$ acts like `anti-friction' and favours the motion of the field climbing up its potential, while the gradient term $\partial V/\partial\psi$ opposes the field's climb, as deduced from inspecting $V(\psi)$ in Figure \ref{fig:quasi-dust-ekpyrotic-potentials}. In fact, while $\psi$ behaves like quasi-dust early on, the field accelerates up its potential, driven by the Hubble term being slightly stronger than the gradient term. 
Eventually, the ekpyrotic energy density becomes comparable to the quasi-dust energy density, and soon enough the ekpyrotic field takes over the dynamics of the background. 
Dominated by the ekpyrotic field rather than the quasi-dust field, the universe contracts at a slower rate.
As a result, the acceleration induced by the Hubble term is no longer larger than the gradient term, and $\psi$ decelerates its motion up its potential. Once the field starts decelerating, the Hubble term remains weaker than the gradient term, perpetuating the deceleration in the motion up the potential. The result is that $\psi$ practically freezes at a positive value of potential energy, with negligible kinetic energy, throughout the ekpyrotic-dominated phases of contraction and ensuing expansion after the bounce. While it is the subdominant field, its behaviour is akin to that of dark energy with an equation of state~$-1$~%
\footnote{As the universe expands, this `frozen' potential energy of $\psi$ does not dilute, unlike the energy density of the dominant ekpyrotic field. Consequently, the subdominant field eventually overtakes the dominant field. For a different choice of parameters than those we constrain this work to, this process repeats cyclically, leading to an alternation in field dominance that drives intriguing dynamics in this cosmological model. However, such an alternation leads to a new bounce every time that the ekpyrotic field rolls across its potential well, compromising the investigation on the propagation of scale-invariant perturbations through a single bounce.}.
This interesting behaviour will be described in further detail in the following subsection.

In light of the unusual dynamics of the quasi-dust field when it becomes subdominant, we set $\ppsiB = 0$ as the initial condition for the momentum of the quasi-dust field. A non-zero quasi-dust momentum at the bounce would anyway be rapidly dampened out by the Hubble term and vanish shortly to both the past and future of the bounce. Given this choice, the quasi-dust energy at the bounce is given simply by its potential energy, that is $\rho_{\psi B} = V(\psi_B)$.

To have a better control of the extent to which the ekpyrotic field dominates over the quasi-dust field throughout the bounce phase, we introduce the parameter
\be \label{eq:fParameter}
    f = \frac{\rho_{\psi B}}{\rho_{\phi B}} \;,
\ee
representing the ratio of the quasi-dust energy density to the ekpyrotic energy density at the bounce. Notably, at the bounce, $\rho_B = \rho_{\phi B} + \rho_{\psi B} =\rhoc$. Since the quasi-dust is set to have vanishing momentum at this point, its field value at the bounce is determined by $f$ via \footnote{We implicitly choose the positive square root as the argument of $\arcsech$ in \eqref{eq:psiB_init}. Since the potential is symmetric about $\psi = 0$, choosing the positive or negative square root does not affect the physics; it remains invariant under a change of sign of $\psi$.} 
\be \label{eq:psiB_init}
\psi_B = \frac{1}{A} \arcsech{ \left[ \sqrt{\frac{2}{1-\epsilon} \frac{f}{f+1}} \, \right] } \;.
\ee
For the numerical computation we set $f = 2.1 \times 10^{-15}$, as we find this value to result in a good fit to current cosmological observations, particularly regarding the amplitude of scalar curvature perturbations.

Lastly, the value of the ekpyrotic momentum at the bounce, $\pphiB$, is determined from the vanishing effective Hamiltonian constraint \eqref{eq:effHam}, and we specify its sign so that the ekpyrotic field is moving away from the bottom of the potential well at the bounce. To evaluate it, we must first discuss our choice of parameters for the ekpyrotic and quasi-dust potentials.

In consideration of previous studies of the matter-ekpyrotic bounce \cite{Li:2020pww,Cai:2013kja,Cai:2014zga}, we set the parameters of the ekpyrotic potential \eqref{eq:EkpyroticPotential} to
\be
    \label{eq:EkpyroticParameters}
    U_0 = 0.75, \quad \alpha = 30, \quad \beta = 5 \;,
\ee
resulting in a minimum value for the ekpyrotic potential of $U_{min} = -0.956$ at $\phi_{min} = -0.00894$. 

Furthermore, in the case of the matter-bounce scenario studied in \cite{Wilson-Ewing:2012lmx}, featuring only one scalar quasi-dust field in a LQC setting, it was established that a slightly-negative equation of state $P=-\epsilon \, \rho$ during the contraction phase would be able to account for the red-tilt in the scalar curvature power spectrum. In that case, the relation between the scalar spectral index and the equation of state parameter was found to be $n_s=1-12\,\epsilon$ to first order in $\epsilon$. Thus, with the purpose of matching current observational constraints on the scalar spectral index (as will be further discussed at the end of Section \ref{sec:scalar_perturbations}), we set the value of the parameter of the quasi-dust potential to
\be \label{eq:epsilon}
    \epsilon = 0.0029 \, .
\ee
In turn, from \eqref{eq:psiB_init} we then obtain $\psi_B=3.98$ .

After having specified the choices of potential parameters, from the vanishing of the Hamiltonian constraint \eqref{eq:effHam} we obtain $\pphiB=1.52$. We have now completely specified the initial conditions at the bounce and the values of the potential parameters. These are collected in Table \ref{tab:initial_conditions} and Table \ref{tab:parameters} for the convenience of the reader. We now proceed with the analysis of the numerical solution of the background dynamics resulting from the specified initial conditions and parameters.


\renewcommand{\arraystretch}{1.5}  

\begin{table}[htb]
    \centering
    \begin{tabular}{
        |
        @{\hspace{3mm}}c@{\hspace{3mm}} 
        ||
        @{\hspace{5mm}}c@{\hspace{5mm}} 
        |
        @{\hspace{5mm}}c@{\hspace{5mm}} 
        |
        @{\hspace{4.8mm}}c@{\hspace{4.8mm}} 
        |
        @{\hspace{4.7mm}}c@{\hspace{4.7mm}} 
        |
        @{\hspace{3.9mm}}c@{\hspace{3.9mm}} 
        |
        @{\hspace{2.5mm}}c@{\hspace{2.5mm}} 
        |
        @{\hspace{4mm}}c@{\hspace{4mm}} 
        |
    }
        \hline
        \textbf{Initial Condition} & $t_B$ & $v_B$ & $b_B$ & $\phi_B$ & $p_{\phi B}$ & $f$ & $p_{\psi B}$ \\
        \hline
        \textbf{Value} & $0$ & $1$ & $\raisebox{0.7ex}{$\dfrac{\pi}{2\lambda}$}$ & $0$ & $1.52$ & $2.1 \times 10^{-15}$ & $0$ \\
        \hline
    \end{tabular}
    \caption{Summary of initial conditions at the bounce used to solve numerically the background dynamics.}
    \label{tab:initial_conditions}
\end{table}

\begin{table}[htb]
    \centering
    \begin{tabular}{
        |
        @{\hspace{4mm}}c@{\hspace{4mm}} 
        ||
        @{\hspace{5mm}}c@{\hspace{5mm}} 
        |
        @{\hspace{6mm}}c@{\hspace{6mm}} 
        |
        @{\hspace{6.5mm}}c@{\hspace{6.5mm}} 
        |
        @{\hspace{4mm}}c@{\hspace{4mm}} 
        |
    }
        \hline
        \textbf{Parameter} & $U_0$ & $\alpha$ & $\beta$ & $\epsilon$ \\
        \hline
        \textbf{Value} & $0.75$ & $30$ & $5$ & $0.0029$ \\
        \hline
    \end{tabular}
    \caption{Summary of values of parameters for ekpyrotic and quasi-dust potentials used to solve numerically the background dynamics.}
    \label{tab:parameters}
\end{table}


\subsubsection{Evolution of the background}

\begin{figure}[t]
    \centering
    \makebox[\linewidth]{%
        \includegraphics[width=0.5\textwidth]{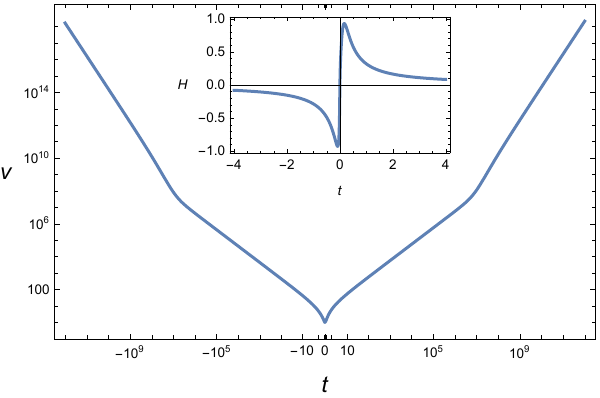}
        \hspace{12pt}
        \includegraphics[width=0.53\textwidth]{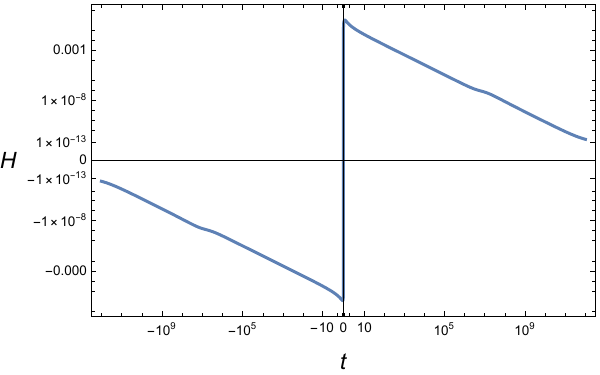}
    }
    \caption{(Left) Evolution of the volume over time. The inset of the Hubble parameter highlights that the bounce, at $t_B=0$, is non-singular. The contraction is first dominated by quasi-dust (steeper descent) and then the ekpyrotic field (shallower descent). The bounce phase takes places in the vicinity of $t_B=0$ (non-singular dip). The expansion is first dominated by the ekpyrotic field (shallower ascent) and then by quasi-dust (steeper ascent).
    (Right) Evolution of the Hubble parameter over time. The horizontal axes of the main panels represent time on a signed-logarithmic scale, while the vertical axes are logarithmic for the left panel and signed-logarithmic for the right panel. For the inset, all axes are linear.
}
    \label{fig:volume-hubble-comparison}
\end{figure}

\begin{figure}[t]
    \centering
    \makebox[\linewidth]{%
        \hspace{-6pt}
        \includegraphics[width=0.51\textwidth]{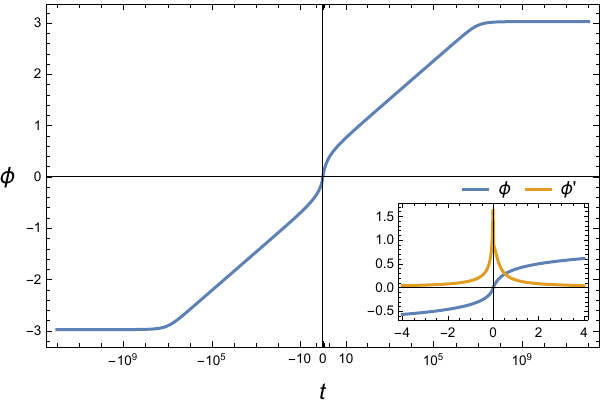}
        \hspace{12pt}
        \includegraphics[width=0.51\textwidth]{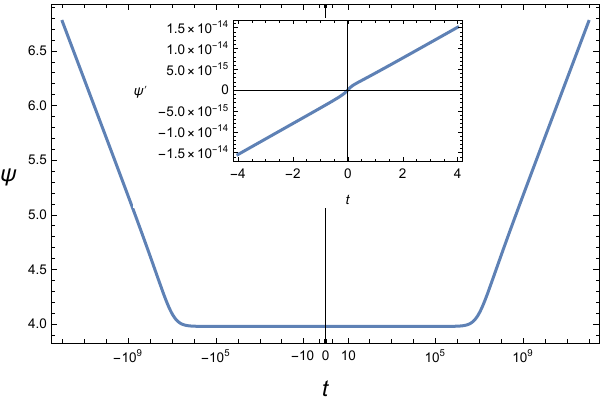}
    }
    \caption{(Left) Evolution of the ekpyrotic scalar field over time, with an inset highlighting its behaviour near the bounce.
    (Right) Evolution of the quasi-dust scalar field over time, with an inset highlighting the behaviour of its derivative with respect to time near the bounce.
    The horizontal axes of the main panels represent time on a signed-logarithmic scale, while the vertical axes are linear. For the insets, all axes are linear.}
    \label{fig:phi-psi-evolution}
\end{figure}

\begin{figure}[t]
    \centering
    \makebox[\linewidth][c]{%
        \hspace{-8pt}
        \includegraphics[width=0.56\textwidth]{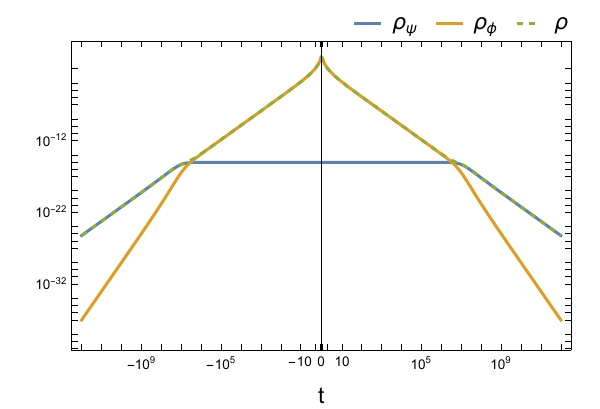}
        \hspace{-12pt}
        \includegraphics[width=0.56\textwidth]{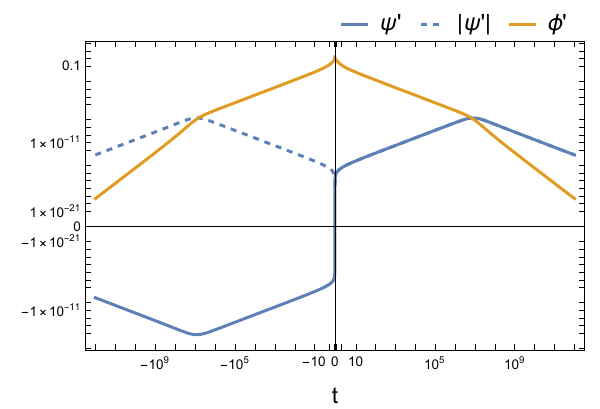}
    }
    \caption{(Left) Energy densities of the quasi-dust and ekpyrotic scalar fields, $\rho_\psi$ (blue) and $\rho_\phi$ (orange) respectively, as well as the total energy density $\rho$ (green dashed). 
    The energy densities are equal at $t_{eq1} = -3.64 \times 10^6$ before the bounce and $t_{eq2} = 3.64 \times 10^6$ after.
    (Right) Field speeds $\dot{\psi}$ (blue) and $\dot{\phi}$ (orange). The field speeds are equal in magnitude at $t = -6.98 \times 10^6$ before the bounce and $t = 6.98 \times 10^6$ after. 
    The horizontal axes of both plots represent time on a signed-logarithmic scale. The vertical axis of the left plot is logarithmic while it is signed-logarithmic for the right plot.
    Notably, the field speeds' magnitudes overtake domination at slightly different times than the fields' total energy densities.}
    \label{fig:energy-densities-and-field-speeds}
\end{figure}

With the choice of parameters and initial conditions as collected in Tables \ref{tab:initial_conditions} and \ref{tab:parameters}, we simulate the dynamics of the background by numerically solving the system of differential equations \eqref{eq:bdot}–\eqref{eq:ppsidot}. We find that the evolution of the volume undergoes a smooth, non-singular bounce at $t_B=0$, throughout which the Hubble parameter remains finite, as can be seen in Figure \ref{fig:volume-hubble-comparison}.
The evolution of the volume is slightly asymmetric with respect to the bounce time. This asymmetry is induced by the ekpyrotic potential, which is asymmetric due to the parameter choice $\beta=5$. This is the only asymmetric element introduced in this model. The rate of change in volume is slower during the phase when the ekpyrotic field is in the right, steeper branch of its potential compared to when the field is in the left, shallower branch. In other words, the universe contracts at a faster rate than it later expands at after the bounce while dominated by the ekpyrotic field. However, the difference in the rate of change of volume and the overall asymmetry it induces in the background is very small and can be hard to discern from the plots in Figures \ref{fig:volume-hubble-comparison}, \ref{fig:phi-psi-evolution} and the left panel in Figure \ref{fig:energy-densities-and-field-speeds}.
It is worth remarking that the evolution of the background remains smooth throughout the whole time span. Even though there are times when derivatives can grow very large quickly, they always remain finite.


From Figure \ref{fig:energy-densities-and-field-speeds} it can be seen that the universe undergoes different phases of domination. The time span we explore begins with a quasi-dust-dominated phase of contraction. After the time $t_{eq1} = -3.64 \times 10^6$, the universe transitions into a phase of ekpyrotic-dominated contraction that continues until the bounce phase in the vicinity of $t_B=0$. The universe remains ekpyrotic-dominated through the bounce regime, when quantum gravity effects become important. The bounce then gives way to a phase of ekpyrotic-dominated expansion. Finally, the dilution of the ekpyrotic energy due to the expansion results in a transition to a phase of quasi-dust-dominated expansion after the transition time $t_{eq2} = 3.64 \times 10^6$.
\footnote{Note that although $t_{eq1}$ and $t_{eq2}$ have the same absolute value rounded to 3 significant figures, they actually differ when rounded to 8 significant figures, with the difference being $\mathcal{O}(0.1)$. The changes in domination happen at different times before and after the bounce as a consequence of the asymmetry in the evolution of the background induced by the choice $\beta=5$. If instead $\beta=1$ was chosen while keeping the initial conditions the same, the background evolution would be symmetric in time around the bounce, and  $t_{eq1}$ and $t_{eq2}$ would simply be opposite values indeed, but this is not the case in this work.}
The bounce takes place during a period of ekpyrotic domination, as seen in Figure \ref{fig:energy-densities-and-field-speeds}, which aligns with the premise that the ekpyrotic field's energy density eventually dominates all other contributions as the volume decreases.

Early on in the quasi-dust-dominated phase of contraction, the dynamics of the background is approximately that of a perfect fluid with a constant equation of state $w_T \approx-\epsilon$.
The effect of the ekpyrotic field's growing energy density is small but non-vanishing. 
It induces a small deviation in the Hubble term, which at this time is the term that drives the motion of the dominant quasi-dust field. 
The acceleration resulting from the Hubble term is smaller than it need be for the quasi-dust field to remain a perfect fluid with equation of state parameter $-\epsilon$.
As a result, initially the total equation of state turns out to instead be $w_T(t_i)=-0.002903$, which is only $0.09\%$ below $-\epsilon$ and approximately constant throughout $t\in [-10^{12},-10^{10}]$.
Such a total equation of state effectively captures a cosmological setting dominated by pressureless dust but also a small, yet significant, presence of dark energy with equation of state $-1$. As illustrated in Figure \ref{fig:phi-psi-evolution}, the quasi-dust field starts at a positive value $\psi(t_i) = 6.77$ and decreases as the universe contracts, climbing up its potential while it is the dominant field.


The ekpyrotic field begins with a value $\phi(t_i) = -2.98$ as seen in Figure \ref{fig:phi-psi-evolution}, which is far in the left branch of its potential. The potential's slope is practically flat there, inducing only a negligible acceleration on the field.
Nonetheless, the universe being dominated by the quasi-dust field contracts at a faster rate than it would if it were ekpyrotic-dominated, leading to a stronger acceleration induced by the Hubble term on the ekpyrotic field as it rolls along the part of its potential with a nearly flat slope towards the potential well.
During this stage, the motion of $\phi$ is dominated by its kinetic term while its potential is negligible. Thus, the ekpyrotic equation of state parameter is $w_\phi \approx 1$ already at $t_i$. This corresponds to a period of fast-roll contraction. It is therefore already competing with the growth of anisotropies at this stage, and not only near the bounce.

As the universe continues to contract under quasi-dust domination, the ekpyrotic field continues to roll down its potential, with its speed steadily increasing due to the Hubble term, which in turn implies a continuous rise in its kinetic energy. During this phase, the energy density of the ekpyrotic field grows at a faster rate than that of quasi-dust. At $t_{eq1}$ the ekpyrotic energy density overtakes the quasi-dust energy density, marking the onset of the ekpyrotic-dominated phase of contraction leading to the bounce. Beyond this point, as the ekpyrotic field continues its descent along its increasingly-steep potential to larger negative values it will in turn gain kinetic energy.


\begin{figure}[t]
    \centering
    \makebox[\linewidth][c]{%
        \hspace{-10pt}
        \includegraphics[width=0.58\textwidth]{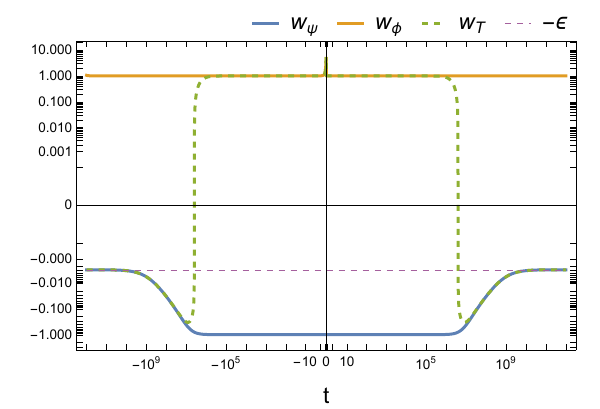}
        \raisebox{4pt}{
            \includegraphics[width=0.52\textwidth]{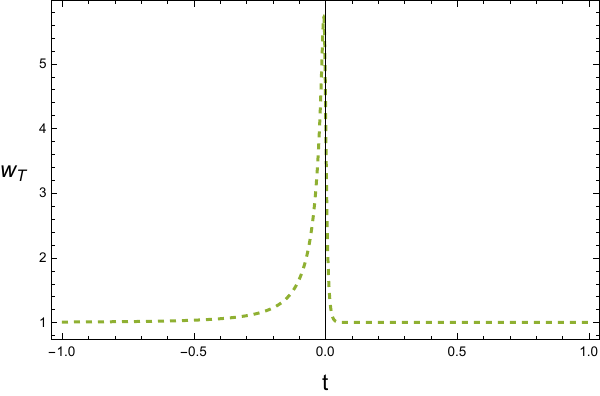}
        }
    }
    \caption{(Left) Equations of state of the quasi-dust and ekpyrotic scalar fields, $w_\psi$ (blue) and $w_\phi$ (orange) respectively, as well as the total equation of state $w_T$ (green dashed). The horizontal axis represents time on a signed-logarithmic scale, while the vertical axis is logarithmic. (Right) Total equation of state $w_T$ focused around the bounce phase, showing the ultra-stiff phase. The maximum value $w_T = 5.72$ takes place at $t=-0.0059$. Both axes are linear.}
    \label{fig:equations-of-state}
\end{figure}


Just before the bounce, the ekpyrotic field starts rolling down the steep potential well. As it rolls down, it quickly gains kinetic energy while dropping in potential energy to more negative values, such that $P_{\phi} > \rho_{\phi}$, resulting in an equation of state parameter larger than $1$. The equation of state parameter peaks at $w_T = 5.72$  immediately before the bounce, when the field is at the minimum of the potential.
As the field traverses the bottom of the potential its total energy density reaches the critical energy density $\rhoc$ (modulo the insignificant contribution from the subdominant quasi-dust energy density) when the field reaches $\phi_B = 0$ and has just initiated its ascent up the right, steeper branch of its potential.
Recall that for the ekpyrotic energy density $\rho_\phi \propto a^{-3(1+w_\phi)}$ and for anisotropic stress-energy $\rho_{anis} \propto a^{-6}$. During the brief period before the bounce where $w_\phi>1$, the ekpyrotic field is expected to ultimately dominate over anisotropic stress-energy as the bounce is approached. In this way, this cosmological scenario avoids evolving towards a BKL instability as the universe contracts.
In fact, the ekpyrotic field's equation of state is $w_\phi \geq 1$ throughout the whole time span explored, as can be seen in Figure \ref{fig:equations-of-state}. 
This implies that $\rho_\phi\geq\rho_{anis}$ always as long as it is the case already at $t_i$.

Interestingly, when the field $\psi$ is not dominant, it does not behave like a quasi-dust fluid. At the last stages of quasi-dust-dominated contraction before $t_{eq1}$, the ekpyrotic energy density becomes relevant even if still subdominant. The ekpyrotic field's growing contribution to the total energy density slows down the rate of contraction, which in turn reduces the acceleration induced by the Hubble term on $\psi$.
Consequently, the reduced Hubble term cannot match the opposing gradient term, dampening the motion of $\psi$ as it climbs up its potential.
As it loses kinetic energy while still gaining potential energy, the field's equation of state parameter starts to drop towards a value of $-1$, characteristic of dark energy. The deceleration of $\psi$ starts shortly before it stops being the dominant component. Hence the total equation of state parameter reflects the negative equation of state of $\psi$ briefly before the transition time. It reaches a minimum value of $w_T=-0.335$ at $t=-8.48 \times 10^6$ as can be seen in Figure \ref{fig:equations-of-state}. Soon after, the ekpyrotic field comes to dominate and the total equation of state transitions to that of the ekpyrotic field $w_T \approx w_\phi = 1$. In contrast, the $\psi$ field continues to decelerate swiftly and freezes at a potential value for which  the Hubble term it experiences offsets the acceleration due to the potential's slope. As a result, during ekpyrotic domination, the field $\psi$ behaves like a constant energy density, akin to dark energy or a cosmological constant with equation of state parameter $w_\psi \approx -1$.

After the bounce, the ekpyrotic field remains kinetically dominated as it climbs up the potential well although it starts to lose kinetic energy. The bounce phase is thus followed by a phase of ekpyrotic-dominated fast-roll expansion. The ekpyrotic energy density dilutes quickly during the expanding phase but the quasi-dust energy density does not since $\psi$ remains frozen. Consequently, the quasi-dust energy density overtakes the ekpyrotic energy density at $t_{eq2}$ and gives way to a phase of quasi-dust dominated expansion.
The transition in domination takes place while the quasi-dust field starts to accelerate due to the reduced Hubble friction not fully offsetting the gradient term anymore, but the speed of the field is not yet adequate for it to behave like a quasi-dust fluid. Hence, the equation of state parameter of the universe can be seen in Figure \ref{fig:equations-of-state} to drop from $1$ to a minimum of $w_T=-0.335$ at $t=8.48 \times 10^6$, and then steadily increase towards a value $w_T(t_f)=-0.002903$ again at the end of the time span, only $0.09\%$ away from the parameter $-\epsilon$. At this stage, the ekpyrotic field is out of the well on the right branch of its potential, climbing upward but at an ever-slower pace due to the stronger Hubble friction induced by the evolving quasi-dust dynamics.

Now that we have thoroughly discussed the intricate dynamics of the background for the specific choice of initial conditions and parameters collected in Tables \ref{tab:initial_conditions} and \ref{tab:parameters}, in the next subsection we will discuss the effects that different choices of parameters and initial conditions have on the background.

\subsubsection{Effects from different parameter choices} \label{sec:Background_Effects-of-different-parameter-choices}

In this subsection, we examine the impact of varying the model's parameters on the evolution of the background within the space of background solutions portrayed earlier. While the qualitative behaviour remains consistent across different parameter choices, each parameter influences specific aspects of the dynamics. For clarity and brevity, we do not present all the corresponding plots of the background solution obtained for each variation, but note that they show qualitatively similar trends, which will be highlighted in this discussion.

The dynamics of the resulting background is primarily influenced by the choice of $f$, which in turn, given the fixed condition $\ppsiB = 0$, determines the value of $\psi_B$ at the bounce.
For our discussion on the evolution of the background we set $f=2.1 \times 10^{-15}$ because we find this value to produce a scalar curvature power amplitude that matches current observations, as we will discuss in the next section. A different choice of $f$ effectively would change the value of the potential energy at which the quasi-dust field freezes during the time it is subdominant. For example, let's consider increasing the value to $f=2.1 \times 10^{-14}$ while keeping all other parameters and initial conditions the same. This amounts to a smaller value $\psi_B=3.71$ at the bounce, hence the quasi-dust-field is frozen at a higher value of its potential throughout its phase of subdominance.
It can be seen from Figure \ref{fig:energy-densities-and-field-speeds} that a higher quasi-dust potential energy at the bounce would shift the transition times between ekpyrotic and quasi-dust domination closer to the bounce. Thus, the duration of overall ekpyrotic domination is reduced and the duration of quasi-dust dominated contraction and expansion are longer within the set time span, resulting in a number of significant consequences. To start with, the values of $\phi$ at $t_i$ and $t_f$ are smaller in magnitude, hence closer to the potential well, because the period during which a stronger Hubble term is induced is longer. Hence, the ekpyrotic field starts from a smaller value $\phi(t_i)=-2.79$, and similarly finishes at a smaller value too.
Since the rate of change of volume is faster during quasi-dust domination, the volume at $t_i$ and $t_f$ is larger too.
Furthermore, the period of time where the total equation of state parameter is close to $-\epsilon$ before the bounce is longer and ends later, closer to the bounce time. The size of the horizon at the transition time is smaller since these times are closer to the bounce scale.
During this period, scales that exit the horizon will acquire a nearly scale-invariant spectrum, as we will discuss in the next section.

Moreover, in order to yield the correct scalar spectral index, the parameter $\epsilon$ is constrained to $0<\epsilon\ll1$. We find that varying the parameter $\epsilon$ within $0 < \epsilon < 0.01$ has very little impact on the background. It only has an appreciable effect on the value of the total equation of state and the rate of change of volume during the quasi-dust dominated phases of contraction and expansion. For example, setting a smaller value $-\epsilon=-0.0020$ while keeping all the other parameters unchanged results in $w_T(t_i) \approx w_\psi(t_i) =-0.002003$ initially. With an equation of state slightly closer to that of dust, the rate of change is slightly slower, but the effect on the initial and final volumes is inappreciable in our simulations. It has no appreciable effect on other background quantities.

The effects of varying the ekpyrotic parameters are most important throughout the phases of ekpyrotic domination. These effects are consistent with those described in \cite{Li:2020pww} due to the fact that when the ekpyrotic field dominates, the evolution of the field and the geometry approximate that of a cosmology with only said field, both in their model and ours alike.

The parameter $U_0$ is related to the depth of the ekpyrotic potential, which in turn affects its slope too. For example, consider the case of increasing $U_0$ while keeping all other parameters and initial conditions fixed. 
The ekpyrotic potential energy at the bounce decreases to a larger negative value, and therefore the ekpyrotic kinetic energy increases since the total energy must still add up to $\rhoc$.
From equation \eqref{eqPressure} we see that therefore the pressure increases, while the total energy remains the same. Thus, the peak of the total equation of state near the bounce, which is dominated by the ekpyrotic equation of state, increases as a result of increasing $U_0$.
This is because as the ekpyrotic kinetic energy is increased during the period it traverses the potential well, the field rolls down and back up the well faster, but takes a longer time to do so due to the deeper well. The range of field values for which $w_T \approx w_\phi > 1$ is larger due to the steeper well.

The parameter $\alpha$ is related to the width of the ekpyrotic potential, which in turn affects its slope too. For example, consider the case of increasing the value of $\alpha$. This makes the potential well narrower, but also steeper. Meanwhile, the slope of the potential away from the well is flatter, and the value of the potential at a given value of $\phi$ is increased. The ekpyrotic field traverses the potential well faster firstly because the well is narrower and secondly because the increased steepness increases the field speed.
As a result, the duration of the phase with equation of state parameter $w_T \approx w_\phi > 1$, namely the  ekpyrotic phase, is shortened. 
Conversely, decreasing the value of $\alpha$ results in a steeper slope for the potential outside the well. This results in a stronger acceleration of the field towards the well whenever it is rolling outside it. Setting $\alpha=26$ or lower while keeping all other parameters and initial conditions the same results in the ekpyrotic field starting off at $t_i$ moving away from the well and later turning around within the chosen time span. This would unnecessarily complicate the study of perturbations through the bounce, and lies outside the scope of this work.

The parameter $\beta$ controls the degree of asymmetry in the ekpyrotic potential between positive and negative field values.
This has a direct, but very subtle impact on the degree of symmetry of the dynamics to the past and to the future of the bounce.
For $\beta=1$ the ekpyrotic potential is symmetric, and therefore the background evolution may be perfectly symmetric too about the bounce time given appropriate initial conditions.
However, since the ekpyrotic potential in this work is postulated at a phenomenological level only, we consider the value of $\beta=5$ as an arbitrary choice representing the fact that the potential is allowed to be asymmetric.
Increasing the value of $\beta$ increases the asymmetry in the ekpyrotic potential.
In terms of field values, the negative branch of the potential remains largely unaffected by increasing $\beta$, while the positive side is squeezed, increasing the slope of the potential and the value of the potential for a given value of $\phi$.
Increasing $\beta$ has a similar effect to that of increasing $\alpha$ on the positive branch of the potential, but has a negligible effect on the negative branch of the potential.

At this stage, we have a thorough account of a numerical solution to the intricate dynamics of the background for a model of LQC with a quasi-dust scalar field and an ekpyrotic scalar field as its matter content. We will be employing the solution obtained with initial conditions and parameters as summarised in Tables \ref{tab:initial_conditions} and \ref{tab:parameters} to study how scalar and tensor curvature perturbations evolve throughout this cosmological setting in the remaining of the paper.

\section{Dynamics of scalar perturbations and their power spectra} 
\label{sec:scalar_perturbations}

In this section, we discuss the scalar power spectra from the quasi-dust-ekpyrotic bounce scenario in LQC. 
We start by laying out the Hamiltonian formalism for scalar perturbations in a setting with two scalar fields. 
Then, we find approximate initial solutions to set in the early, classical regime, where the evolution of the perturbation modes is approximately decoupled.
To study the propagation of perturbations we will employ the effective description of the \emph{dressed metric} approach \cite{Agullo:2012fc,Li:2019qzr}, in which quantum perturbations evolve on a quantum spacetime that is well-approximated by a differential manifold equipped with a dressed metric that captures the effects of quantum gravity for sharply-peaked semiclassical states.
We then solve numerically for the evolution of scalar perturbations and compute their curvature and entropic power spectra.
Finally, we discuss the results and compare them to current observations to assess the validity of the scenario.

\subsection{Formalism for scalar perturbations with two scalar fields}

Following the Hamiltonian formalism developed in \cite{Pascual:2025teu}, for a cosmological setting with its matter sector comprised of two scalar fields, the scalar subspace of linear perturbations in Fourier space is spanned by 
\be
\Gamma^{(1)}_S = \Big\{ \gamma_1, \gamma_2, \delta \phi, \delta \psi\Big\}
\ee
where $\gamma_1$ and $\gamma_2$ are the two scalar components of the linear perturbation of the Fourier transformed spatial metric, and $\delta \phi$ and $\delta \psi$ are the Fourier coefficients of the linear perturbations of the ekpyrotic and quasi-dust fields respectively.
We do not explicitly label each Fourier coefficient with its corresponding comoving wavenumber $k$ to prevent our notation becoming too cluttered, but it should be implicit from context that perturbation variables are coefficients of Fourier modes $k$.
The two physical, gauge-invariant degrees of freedom in $\Gamma^{(1)}_S$ can be chosen to be characterised by the canonical Mukhanov-Sasaki variables
\be \label{eq:Perturbation-definition}
\Qphi = \delta\phi + \frac{3 \pphi}{8 \pi G a \pi_a} \left( \gamma_1 - \frac{1}{3} \gamma_2 \right) 
\quad \text{and} \quad
\Qpsi = \delta\psi + \frac{3 \ppsi}{8 \pi G a \pi_a} \left( \gamma_1 - \frac{1}{3} \gamma_2 \right) \;,
\ee
where $\pi_a$ is the conjugate momentum of the scale factor. 
Moreover, in terms of their Fourier modes, the equations of motion for the Mukhanov-Sasaki variables are given by
\begin{gather} \label{eq:QPhi-EoM}
\ddot\Qphi + 3H\dot\Qphi + \frac{k^2 + \Omega^2_{\phi\phi}}{a^2} \, \Qphi + \frac{\Omega^2_{\phi\psi}}{a^2} \Qpsi = 0 \;, 
\\[6pt] \label{eq:QPsi-EoM}
\ddot\Qpsi + 3H\dot\Qpsi + \frac{k^2 + \Omega^2_{\psi\psi}}{a^2} \, \Qpsi + \frac{\Omega^2_{\phi\psi}}{a^2} \Qphi = 0 \;,
\end{gather}
where 
$k^2=k_i k_j \, \delta^{ij} = a^2 k_i k^i$ denotes the comoving wavenumber squared, and
the different $\Omega^2_{FF'}$ for the subscripts $F,F'\in\{\phi,\psi\}$ all depend only on background quantities and are explicitly given by
\begin{align} 
\label{eq:OmegaPhiPhi}
\Omega^2_{\phi\phi} =& \,
24 \pi G \frac{p_\phi^2}{a^4} 
- 18 \frac{p_\phi^2}{a^6 \pi_a^2} \left( \pphi^2 + \ppsi^2 \right)
- 12\frac{a}{\pi_a} p_\phi \frac{\partial U}{\partial \phi}
+ a^2 \frac{\partial^2 U}{\partial \phi^2} \, , 
\\[6pt] \label{eq:OmegaPhiPsi}
\Omega^2_{\phi\psi} =&  \,
24 \pi G \frac{p_\phi p_\psi}{a^4} 
- 9 \frac{p_\phi p_\psi}{a^6 \pi_a^2} \left( \pphi^2 + \ppsi^2 \right)
- 6\frac{a}{\pi_a} \left( p_\psi \frac{\partial U}{\partial \phi} + p_\phi \frac{\partial V}{\partial \psi} \right) \, , \\[6pt] \label{eq:OmegaPsiPsi}
\Omega^2_{\psi\psi} =& \,
24 \pi G \frac{p_\psi^2}{a^4} 
- 18 \frac{p_\psi^2}{a^6 \pi_a^2} \left( \pphi^2 + \ppsi^2 \right)
- 12\frac{a}{\pi_a} p_\psi \frac{\partial V}{\partial \psi}
+ a^2 \frac{\partial^2 V}{\partial \psi^2} \ .
\end{align}


\begin{figure}[t!]
    \centering
    \includegraphics[width=0.7\textwidth]{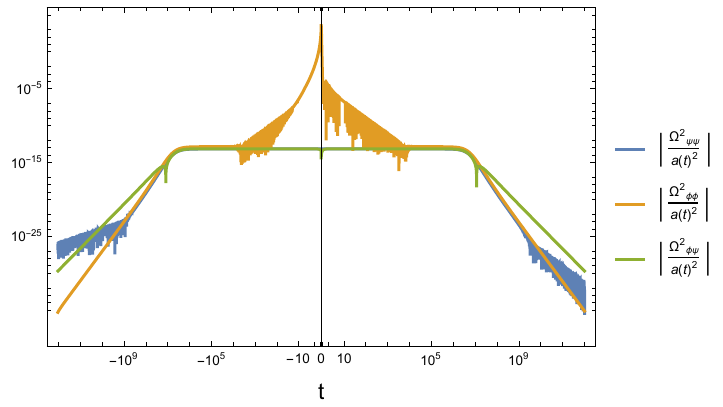}
    \caption{Plots of the factors with $\Omega_{\phi\phi}$, $\Omega_{\psi\psi}$ and $\Omega_{\phi\psi}$ entering the equations of motion \eqref{eq:QPhi-EoM} and \eqref{eq:QPsi-EoM} as functions of time. The horizontal axis represent time on a signed-logarithmic scale, while the vertical axis is logarithmic.}
    \label{fig:Omegas}
\end{figure}


The background quantities in Equations \eqref{eq:OmegaPhiPhi}-\eqref{eq:OmegaPsiPsi} are determined from the solutions of the background dynamics we found in Section \ref{sec:Background_Dynamics}. 
That said, different forms of $\Omega^2_{FF'}$ can be reached based on the exact form of the zeroth-order constraint one considers.
Since the background dynamics obeys the LQC modified Friedmann Equation \eqref{eq:LQCFriedmann} in the effective approach, it is natural to use the effective Hamiltonian constraint \eqref{eq:effHam} to replace $\pi_a$ in the $\Omega^2_{FF'}$ in Equations \eqref{eq:OmegaPhiPhi}-\eqref{eq:OmegaPsiPsi} with the background solution computed, as was suggested in \cite{Li:2019qzr,Li:2020pww}. This amounts to making the replacement
\begin{align} \label{eq:pi2_a_Replacement}
\frac{1}{\pi_a^2} &\rightarrow \frac{16 \pi^2 G^2 \gamma^2 \lambda^2}{9 a^4 \sin^2 (\lambda b)} \, , 
\\[6pt] \label{eq:pi_a_Replacement}
\frac{1}{\pi_a} &\rightarrow \frac{-4 \pi G \gamma \lambda \cos (\lambda b)}{3 a^2 \sin (\lambda b)} \, .
\end{align}
Note that the $\cos(\lambda b)$ factor in Equation \eqref{eq:pi_a_Replacement}
is motivated by the treatment of odd powers of $\pi_a$ in the hybrid approach \cite{Fernandez-Mendez:2013jqa,Gomar:2014faa} , and renders $1/\pi_a$ smooth near the bounce since it behaves practically like a step function across the bounce.

The equations of motion of the scalar perturbation modes are coupled via the last term in each of Equations \eqref{eq:QPhi-EoM}-\eqref{eq:QPsi-EoM} with the coupling factor $\Omega^2_{\phi\psi}$. We will solve this coupled system of differential equations numerically, substituting the background solutions computed in Section \ref{sec:Background_Dynamics} appropriately. The initial states for the perturbations are set at $t_i$, deep in the quasi-dust-dominated phase of contraction. At this time, the matter density and the curvature are far below the Planck scale, such that quantum gravity effects can be safely disregarded. Furthermore, the different $\Omega^2_{FF'}$ terms approximately vanish at $t_i$ due to the vanishing of the potentials and their derivatives, and the system of equations can be solved analytically to find an expression for the initial states for the perturbations.
The initial states for the perturbations will be worked out in the following subsection.

In order to compare our results with observations, we relate the scalar perturbation modes described so far to the comoving adiabatic curvature perturbation $\Rcurv$ and the entropy perturbation $\mathcal{S}$
\cite{Garcia-Bellido:1995hsq, Gordon:2000hv}. The relations are given by
\be \label{eq:R-and-S}
\Rcurv = H \frac{\dot{\phi} \Qphi + \dot{\psi} \Qpsi}{\dot{\phi}^2 + \dot{\psi}^2} 
\quad \text{and} \quad   
\mathcal{S} = H \frac{\dot{\phi} \Qpsi - \dot{\psi} \Qphi}{\dot{\phi}^2 + \dot{\psi}^2} \;,
\ee
and their power spectra are given by
\be \label{eq:PowerSpectra-R-S}
\PSR = \frac{k^3}{2\pi^2} |\mathcal{R}|^2
\quad \text{and} \quad
\PSS = \frac{k^3}{2\pi^2} |\mathcal{S}|^2 \; .
\ee

Furthermore, we define the power spectra of $\Qphi$ and $\Qpsi$ respectively as
\be \label{eq:PowerSpectra-Phi-Psi}
\PSphi = \frac{k^3}{2\pi^2} \left| \Qphi \right|^2
\quad \text{and} \quad
\PSpsi = \frac{k^3}{2\pi^2} \left| \Qpsi \right|^2
\ee
to enable us to compare the contributions coming from each Mukhanov-Sasaki variable to the total curvature power spectra.

Lastly, the scalar spectral index $n_s$ quantifies how $\PSR$ depends on the scale $k$. In our numerical analysis, we evolve a wide range of discrete modes, $k \in [10^{-15},10^{-2}]$, which are uniformly logarithmically spaced with an interval of $\delta (\ln k)=0.230$; this corresponds to sampling 10 points per order of magnitude in $k$. As a result, our outcome is a set of discrete points relating $\PSR$ to $k$ rather than a continuous function with a well-defined derivative. Consequently, we resort to a finite-difference approximation to evaluate the formally defined scalar spectral index, both given by
\begin{equation}
    \begin{aligned} \label{eq:n_s-definition}
        n_s(k) - 1 
        &= \frac{ \text{d} \ln\PSR(k) }{ \text{d} \ln k } 
        \\
        &\approx \frac{ \ln\PSR(k e^{\delta (\ln k)}) - \ln\PSR(k) }{ \delta (\ln k) } \, ,
    \end{aligned}
\end{equation}
and similarly for the so-called running of the scalar spectral index $\alpha_s$, which quantifies the scale dependence of the spectral index $n_s(k)$, we use the formal definition and the finite-difference approximation given by
\begin{equation}
    \begin{aligned} \label{eq:running-alpha_s-definition}
        \alpha_s(k) 
        &= \frac{ \text{d} n_s(k) }{ \text{d} \ln k } 
        \\
        &\approx \frac{ n_s(k e^{\delta (\ln k)}) - n_s(k) }{ \delta (\ln k) } \, .
    \end{aligned}
\end{equation}

\subsection{Initial perturbations during quasi-dust-dominated contraction} \label{sec:Scalar_Initial-perturbations}

Early in the quasi-dust-dominated phase of contraction, the total energy density $\rho$ is far below the critical energy density $\rho_{c}$.
The term accounting for the corrections due to quantum gravity in the LQC Friedmann equation is negligible, and the evolution of the universe is well approximated by the classical Friedmann equation
\be \label{eq:Friedmann-cl}
H^2 = \frac{8 \pi G}{3} \rho \;,
\ee
along with the continuity equation \eqref{eq:Rho-evolution}.
During the quasi-dust-dominated phase of contraction, the cosmology resembles that of a perfect fluid with an equation of state $w_T(t_i) \approx w_\psi(t_i) \approx -\epsilon$, denoted simply as $w$ in this subsection to simplify notation, and a total energy density $\rho \approx \rho_\psi$.
With regards to the evolution of the background geometry, the ekpyrotic field can be neglected early on in this phase.
Solving Equation \eqref{eq:Rho-evolution} then yields
\be \label{eq:Rho(a)}
\rho (a) = \rho_i {a_i}^{3(1+w)} a^{-3(1+w)} \;.
\ee
where $a_i = a(t_i)$ is the value of the scale factor at the initial time $t_i$ in the time span explored, and $\rho_i=\rho(a_i)$ is the total energy density at $t_i$.
We then solve Equation \eqref{eq:Friedmann-cl} using \eqref{eq:Rho(a)} to find the scale factor as a function of proper time $t$
\be
a(t) = a_i \Big[ 6 \pi G (1+w)^2 \rho_i (t - t_0)^2 \Big]^{\frac{1}{3(1+w)}}
\ee
where $t_0$ is a constant of integration
(which would correspond to the classical big-crunch singularity in a classical cosmological setting with no corrections from LQC).
This relation can be rewritten in terms of $t_i$ as
\be \label{eq:a(t)Init}
a(t) = a_i \Big[ 1 - \sqrt{6 \pi G \rho_i} (1+w) (t - t_i) \Big]^\frac{2}{3(1+w)} 
\ee
which just implies that
$t_0 = t_i - \dfrac{1}{\sqrt{6 \pi G \rho_i} (1+w)} $
.
Moreover, the scale factor in terms of conformal time $\eta$, related via $dt=a\,d\eta$, is given by%
\footnote{The functions $a(t)$ and $a(\eta)$ are of course different by their mathematical definitions, but are both represented as $a$ since they both have the same physical interpretation as the scale factor.
}
\begin{align}
a(\eta) 
&= \left[ \frac{2}{3} \pi G \rho_i {a_i}^{3(1+w)} (1+3w)^2 \right]^\frac{1}{1+3w} 
\left[ - (\eta - \eta_i) + \frac{3}{(1+3w) \sqrt{6 \pi G \rho_i} \, a_i} \right]^\frac{2}{1+3w} \\
&= \left[ \frac{2}{3} \pi G \rho_i {a_i}^{3(1+w)} (1+3w)^2 \right]^\frac{1}{1+3w}
\big( - \eta + \eta_0 \big)^\frac{2}{1+3w}
\label{eq:a(eta)Init}
\end{align}
where $\eta_i$ is defined via $a_i = a(t_i) = a(t(\eta_i))$, and
\be \label{eq:eta_0}
\eta_0 = \eta_i + \frac{3}{(1+3w) \sqrt{6 \pi G \rho_i} \, a_i}
\ee
is the conformal time of the would-be classical big-crunch, such that $t_0 = t(\eta_0)$. In our analysis we choose $\eta(t=0)=0$ without loss of generality, and find a numerical solution for $\eta(t)$ which is invertible. In turn, the value $\eta_0=679$ is determined.

During the quasi-dust-dominated contracting phase, both the ekpyrotic and quasi-dust potentials, and their respective derivatives, can be seen to approximately vanish. Consequently, using the vanishing effective Hamiltonian constraint \eqref{eq:effHam} we see that the different $\Omega^2_{FF'}$ terms approximately vanish at $t_i$ due to the vanishing of the potentials.
Therefore, initially the system of equations describing the evolution of the Mukhanov-Sasaki variable is approximately decoupled. Denoting the variables collectively as $\QF \in \{\Qphi,\Qpsi\}$, their equations of motion simplify to
\be \label{eq:QF-EoM-Initially}
\ddot{\QF} + 3H\dot\QF + \frac{k^2}{a^2} \, \QF = 0 \;.
\ee

The solution to Equation \eqref{eq:QF-EoM-Initially} is most easily worked out in terms of the canonical variables $\VF = a \QF$ in conformal time. The perturbation equation takes the form
\be \label{eq:vF-EoM-a''a}
{\VF}'' + \left( k^2 - \frac{a''}{a} \right) \VF = 0
\ee
where the prime denotes differentiation with respect to conformal time $\eta$. Substituting the scale factor \eqref{eq:a(eta)Init} into \eqref{eq:vF-EoM-a''a}, we obtain
\be \label{eq:vF-EoM-w}
{\VF}'' + \left( k^2 - \frac{1-3w}{(1+3w)^2} \frac{2}{(\eta-\eta_0)^2} \right) \VF = 0 \: .
\ee
Thus, each perturbation mode satisfies a harmonic oscillator equation of motion with a time-dependent mass.

To solve this equation, we first rewrite it in terms of the auxiliary variable $\mathcal{B}= \VF / \sqrt{-(\eta - \eta_0)}$, and then rescale the time variable by a factor of $k$, which transforms the equation into the Bessel differential equation for $\mathcal{B}$ in terms of $k(\eta-\eta_0)$. It's general solution, rewritten back in terms of $\VF$, is found to be
\be \label{eq:vF-solution-general}
\VF = \sqrt{-(\eta - \eta_0)} \left[ A_1 H_\mu^{(1)}(-k (\eta - \eta_0)) + A_2 H_\mu^{(2)}(-k (\eta - \eta_0)) \right]
\ee
where $H_\mu^{(1)}$ and $H_\mu^{(2)}$ are the first and second Hankel functions of order $\mu$ given by
\begin{align} \label{eq:Hankel-order}
\mu & = \sqrt{2 \left( \frac{1-3w}{(1+3w)^2} \right) + \frac{1}{4} } 
\\ \label{eq:Hankel-second-order}
& \approx \frac{3}{2} - 6w + 18w^2 + \mathcal{O}(w^3)
\end{align}
and $A_1$ and $A_2$ are constants to be determined by initial conditions. Note that there is a minus sign in front of $\eta$ in the solution because this solution holds early in the contracting branch, when $\eta<0$.
The Hankel functions are well-behaved in the asymptotic limit for large argument \cite{Arfken:2013}, such that in the limit $\eta \rightarrow - \infty$, \eqref{eq:vF-solution-general} becomes (up to an irrelevant global phase)
\be
\lim_{\eta \rightarrow - \infty} \VF = \sqrt{\frac{2}{\pi k}} \left( A_1 e^{-i k \eta} + A_2 e^{i k \eta} \right) \quad .
\ee

The Wronskian condition \cite{Baumann:2022book}
\be
\VF({\VF}^*)' - ({\VF}^*){\VF}' = i 
\ee
is used to normalise the initial perturbation states given they are approximately decoupled. However, note that this condition will not be preserved throughout their evolution when the coupling between the perturbation variables is relevant.
We impose Bunch-Davies vacuum initial conditions
\be
\lim_{\eta \rightarrow - \infty} \VF = \frac{1}{\sqrt{2 k}} e^{-i k \eta}
\ee
by setting the values of the integration constants to
$ A_1 = \sqrt{\pi/4} $ and $ A_2 = 0 $ ,
such that the solution to \eqref{eq:vF-EoM-w} becomes
\be \label{eq:vF-solution-specific}
\VF = \sqrt{-\frac{\pi}{4} (\eta - \eta_0)} \; H_\mu^{(1)}(-k (\eta - \eta_0)) \;.
\ee
In turn, this implies that the expression for initial states for the scalar perturbations set at $\eta=\eta_i=\eta(t_i)$ is given by
\be \label{eq:QF-solution-initially}
\QF = \frac{1}{2a} \sqrt{-\pi (\eta - \eta_0)} \; H_\mu^{(1)} (-k(\eta - \eta_0)) \;,
\ee
and the initial speed, necessary for the numerical computation of their evolution, is found to be
\be \label{eq:QFdot-solution-initially}
\begin{split}
\dot{\QF} = 
&-\frac{1}{2a} 
\left[
H \sqrt{-\pi (\eta - \eta_0)} 
+
\frac{1}{2a} \sqrt{-\frac{\pi}{\eta - \eta_0}}
 \; \right]
H_\mu^{(1)} (-k(\eta - \eta_0))
\\
&-\frac{k}{4a^2} \sqrt{-\pi (\eta - \eta_0)} 
\left( H_{\mu-1}^{(1)} (-k(\eta - \eta_0)) - H_{\mu+1}^{(1)} (-k(\eta - \eta_0)) \right) \, .
\end{split}
\ee

At this point we can turn our attention to super-horizon modes given by \eqref{eq:QF-solution-initially} in the limit $-k (\eta - \eta_0) \ll 1$.
Recall the small-argument expansion of the first Hankel function, given by \cite{Arfken:2013}
\be \label{eq:Small-argument-Hankel}
H_\nu^{(1)}(x) \approx -i \, \frac{\Gamma(\nu)}{\pi} \left( \frac{2}{x} \right)^\nu \;.
\ee
It follows that super-horizon modes given by \eqref{eq:QF-solution-initially} in the quasi-dust-dominated phase behave as
\be
\lim_{-k (\eta - \eta_0) \ll 1} \QF \sim k^{-\mu} \;.
\ee
During the initial quasi-dust-dominated phase, when $\dot{\psi} \gg \dot{\phi}$, we can deduce that $\PSR \approx \PSpsi$ by means of the definitions of the power spectra \eqref{eq:PowerSpectra-R-S} and \eqref{eq:PowerSpectra-Phi-Psi}. 
Therefore, recalling the definition of the scalar spectral index \eqref{eq:n_s-definition}, we find that super-horizon modes already exhibit a red-tilted power spectrum characterised by 
\be \label{eq:n_s-initially}
n_s = 4 - 2\mu = 1 + 12w -36w^2
\ee
to second order in $w$ after using Equation \eqref{eq:Hankel-second-order}.
Modes that exit the horizon during this phase acquire a red-tilted power spectrum produced by the slightly-negative effective equation of state parameter $w$, as was similarly found in \cite{Wilson-Ewing:2012lmx}.
Naturally, during the initial phase, $ w \approx -\epsilon $, and $\epsilon$ was selected in this work to ensure that the numerical solution for the perturbations aligns closely with current observations of the CMB.
To second order, the expression for the scalar spectral index in terms of the quasi-dust parameter $\epsilon$ becomes
\be \label{eq:n_s-epsilon-initially}
n_s = 4 - 2\mu = 1 - 12\epsilon -36\epsilon^2
\ee

Of course, the perturbations still need to evolve through the ekpyrotic-dominated period. We need to assess whether the red-tilted scale-invariance is preserved throughout the the ekpyrotic-dominated period, including the bouncing regime where quantum gravity effects become significant. This is explored in the next subsection, supported by detailed numerical results for the scalar power spectrum.

\subsection{Numerical results for scalar power spectra} \label{sec:Scalar_Numerical-results}


\begin{figure}[t!]
    \hspace*{-34pt}
    \includegraphics[width=1.1\linewidth]{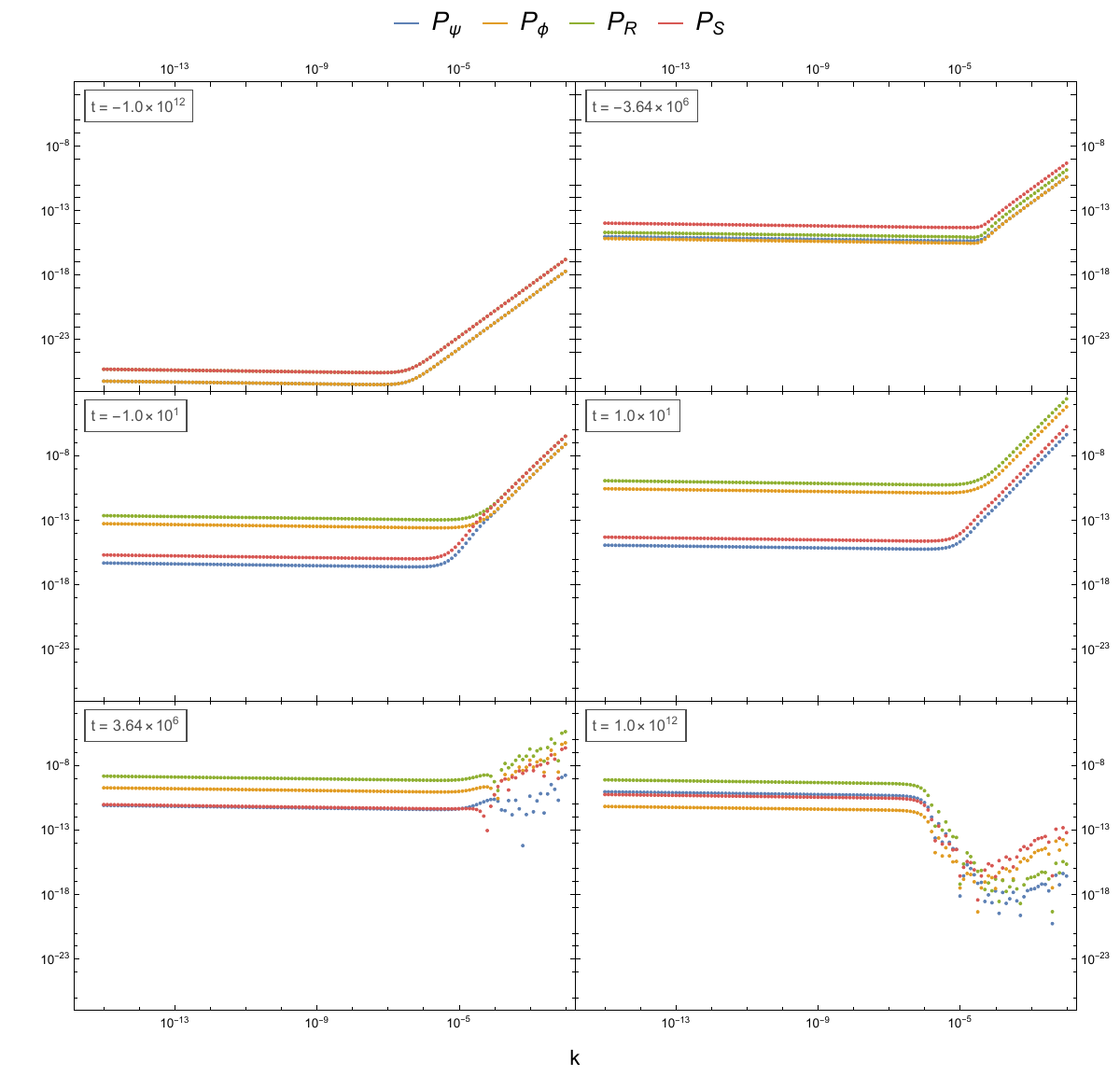}
    \caption{Evolution of the power spectra as functions of $k$ for: curvature perturbations ($\PSR$, green points), entropy perturbations ($\PSS$, red points), ekpyrotic perturbations ($\PSphi$, blue points) and quasi-dust perturbations ($\PSpsi$, orange points) evaluated at different times in chronological order: $t_i = -10^{12}$, $t_{eq1} = -3.64 \times 10^6$, $t = -10$, $t = 10$, $t_{eq2} = 3.64 \times 10^6$, and $t_f = 10^{12}$. Each panel shows the different scalar power spectra evaluated for a discrete sample of uniformly logarithmically spaced comoving wavenumbers $k \in [ 10^{-15} , 10^{-2} ]$ at the labelled time. The vertical logarithmic scales are the same in all panels to facilitate comparison.}
    \label{fig:power-spectra-evolution}
\end{figure}


\subsubsection{Evolution over time of the Power Spectra for different \texorpdfstring{$k$}{}}

The evolution of the amplitude of the different scalar power spectra for a range of values of $k$, namely $k \in [ 10^{-15} , 10^{-2} ]$, is depicted in Figure \ref{fig:power-spectra-evolution}.
From the definition of the comoving curvature and entropic perturbations, Equation \eqref{eq:R-and-S}, we infer that during phases when one field's speed is far larger than the other, $\Rcurv$ is dictated by the perturbation of the field with the larger speed and therefore larger kinetic energy, while $\mathcal{S}$ is dictated by the perturbation of the field with smaller speed. In the background we are considering, the field that dominates the total energy, and therefore the geometrical dynamics, generally also dominates the kinetic energy, thus dictating curvature perturbations during that phase.

Initially, both scalar perturbations $\Qphi$ and $\Qpsi$ exhibit identical behaviour, such that $\PSphi$ and $\PSpsi$ appear practically overlaid. For modes that have exited the horizon, $k \lesssim 10^{-7}$, the power spectrum is nearly scale-invariant, with a slight red-tilt favouring smaller $k$. In contrast, modes still within the horizon, $k \gtrsim 10^{-6}$, display a blue spectrum favouring larger $k$.

Throughout the quasi-dust-dominated phase of contraction, more modes exit the horizon and the red-tilted regime of the power spectrum enlarges to $k \lesssim 10^{-5}$.
The amplitude of the red-tilted regime grows steadily as $\PSR \propto |t|^{-2+4w}$
\footnote{
During the quasi-dust dominated phase of contraction, with equation of state parameter $w$, we find $H=\frac{2}{3(1+w)}|t|^{-1}$ from the scale factor \eqref{eq:a(t)Init}, such that $\dot{\psi} \propto |t|^{-1}$ follows from the Klein-Gordon equation \eqref{eq:K-Geq}. In turn it then follows that for super-horizon modes in this phase, $\Qpsi \propto |t|^{-1+2w}$ to first order in $w$. Therefore, $\PSR \propto (\frac{H}{\dot{\psi}})^2|\Qpsi|^2 \propto |t|^{-2+4w}.$
}
during this phase. 
Both $\Qphi$ and $\Qpsi$ evolve with the same power law since their evolution is approximately decoupled throughout this phase. Notably, the curvature perturbation is dictated by $\Qpsi$ since the quasi-dust field speed is larger in magnitude than the ekpyrotic field speed.
As $t_{eq1}$ is approached, the perturbation variables $\Qphi$ and $\Qpsi$ begin to evolve differently due to the distinct dynamics of the fields.
Shortly after the transition in domination, the ekpyrotic field speed overtakes the quasi-dust field speed, such that the curvature perturbation becomes dictated by the $\Qphi$ term thenafter.

As the system approaches the bounce, at an arbitrarily chosen time $t = -10$, the curvature perturbation $\Rcurv$ is dominated by $\Qphi$.
The amplitude of their red-tilted regime has overall grown up to this point, albeit at a far slower rate compared to the growth undergone during the quasi-dust-dominated phase before $t_{eq1}$.
As the horizon shrinks during the contraction of the universe, curvature modes around $k \sim 10^{-5}$ which were previously outside the horizon have recently entered it, excluding them from the red-tilted regime at this time.
Interestingly, the red-tilted regime of $\PSphi$ extends up to $k \lesssim 10^{-5}$, while for $\PSpsi$ it only extends to $k \lesssim 10^{-6}$. 
The difference arises from the fact that their evolution equations have a different dependence on the background variables, with each perturbation variable having a stronger dependence on its corresponding field than on the other field. As the fields' evolution is completely different, the perturbations inherit a distinct evolution in turn.
The red-tilted regime of $\PSpsi$ has overall slightly decreased since $t_{eq1}$.
For both $\PSphi$ and $\PSpsi$, thus also for $\PSR$,
the blue-regime modes have grown, resulting in a steeper scale dependence proportional to $k^{\,2.65}$, further favouring modes with larger $k$.

After the bounce, at an arbitrarily chosen time $t = 10$, the red-tilted regimes of all perturbation variables extend to the same ranges of $k$ as they respectively did before the bounce. 
Within the red-tilted regimes, $\PSpsi$ is amplified by 1 order of magnitude, while $\PSphi$ is amplified by 3 orders of magnitude through the bounce phase. The large amplification of $\PSphi$ in turn results in a similarly large amplification of the curvature perturbation. The blue regime continues to exhibit the steep scale dependence of $k^{\,2.65}$.

By $t_{eq2}$ in the expanding phase, when $\dot{\phi}\Qphi$ and $\dot{\psi}\Qpsi$ are comparable, the red-tilted regime of $\PSphi$, and consequently of $\PSR$ too, have grown modestly by roughly 1 order of magnitude. In contrast, the red-tilted regime of $\PSpsi$ has undergone a more substantial increase, growing by 4 orders of magnitude due to the amplification of $\Qpsi$ as the phase of quasi-dust-dominated expansion is approached. Meanwhile, the blue regimes of all perturbation variables have evolved into an oscillatory behaviour.

Finally, at $t_f$, when $\dot{\psi}\Qpsi$ dominates $\Rcurv$, the decrease in $\PSphi$ drives a corresponding increase in $\PSpsi$ of similar magnitude, resulting in only a slight overall growth of $\PSR$. Notably, the red-tilted regime of $\PSR$, for $k \lesssim 10^{-7}$, grows only briefly after $t_{eq2}$ and then becomes practically frozen in amplitude. Modes with $k \gtrsim 10^{-7}$, which reenter the horizon in the expanding phase, exhibit an oscillatory behaviour with varying amplitudes.

To illustrate the time evolution of different modes, we focus on two representative examples: 
$k=10^{-10}$, which lies in the red-tilted regime, and $k=10^{-2}$, which becomes oscillatory with a decreasing average amplitude early in the expansion phase (i.e., for $t \gtrsim 10^3$).
Figures \ref{fig:kSuper-Scalar-tEvolution} and \ref{fig:kSub-Scalar-tEvolution} show how $\PSphi$, $\PSpsi$, $\PSR$ and $\PSS$ evolve over time for these modes. The evolution over time of modes $k < 10^{-8}$ is qualitatively identical to that of the representative red-tilted mode $k=10^{-10}$, including the modes that correspond to scales of interest for current observations. Similarly, modes $10^{-4} < k \lesssim 1 $ evolve qualitatively the same as the representative oscillatory mode $k=10^{-2}$, with $ k \gtrsim 1 $ being trans-planckian modes which we are not concerned with.

\subsubsection{Evolution over time of the Power Spectra for \texorpdfstring{$k=10^{-10}$}{}}


\begin{figure}[t!]
    \centering
    \makebox[\linewidth]{%
        \includegraphics[width=0.57\textwidth]{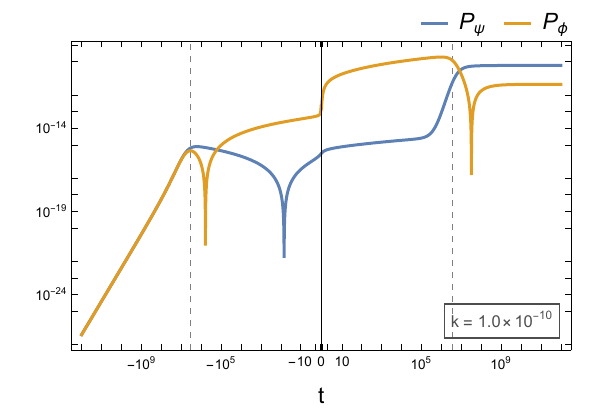}
        \hspace{-20pt}
        \includegraphics[width=0.57\textwidth]{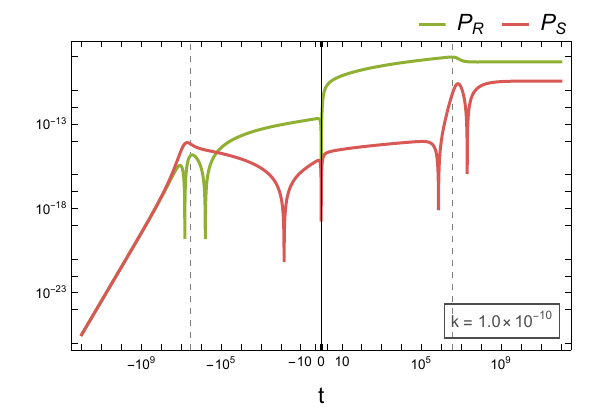}
    }
    \caption{Evolution over time of: (left) $\PSphi$ and $\PSpsi$, and (right) $\PSR$ and $\PSS$, for mode $k=10^{-10}$ within the red-tilted regime. The vertical dashed lines mark the times of equal energy densities $t_{eq1} = -3.64 \times 10^6$ and $t_{eq2} = 3.64 \times 10^6$. The horizontal axes represent time on a signed-logarithmic scale, while the vertical axes are logarithmic. See appendix \ref{app:real-imaginary-components} for further detail regarding the non-vanishing dips of $\PSphi$ at $t=-6.4 \times 10^5$ and $t=3.3 \times 10^7$, and $\PSpsi$ at $t=-71$.}
    \label{fig:kSuper-Scalar-tEvolution}
\end{figure}


We will first consider the evolution of the different power spectra for the mode $k=10^{-10}$.
This mode is outside the horizon for most of the time span explored except for a brief period around the bounce phase, during which the horizon grows to infinity due to the vanishing Hubble parameter at the bounce. This mode lies well within the regime with a red-tilted curvature power spectrum.
As can be seen in Figure \ref{fig:kSuper-Scalar-tEvolution} depicting their evolution, all four power spectra for $k=10^{-10}$ undergo dips and rebounds in their amplitude at different times and for different reasons throughout the time span explored.

The first significant dip and rebound in $\PSR$ appears near $t = -6.8 \times 10^6$ around the time the ekpyrotic field speed overtakes the quasi-dust field speed, just before $t_{eq1}$ in the contracting branch.
Until then, $Q^\phi$ and $Q^\psi$ have been evolving in the same way under equation \eqref{eq:QF-EoM-Initially} and from the same initial quantum vacuum state, so they remain nearly equal in magnitude. However, at this point $|\dot{\phi}|$ overtakes $|\dot{\psi}|$ in magnitude, while $\dot{\psi} < 0$ but $\dot{\phi} > 0$.
\footnote{While $\dot{\phi} > 0$ for all $t \in [t_i, t_f]$, $\dot{\psi} < 0$ for $t \in [t_i, 0]$ and $\dot{\psi} > 0$ for $t \in [0, t_f]$. This can be seen in the right panel of Figure \ref{fig:energy-densities-and-field-speeds}.} 
Consequently, the two terms in $\Rcurv$, namely $\dot\phi\Qphi$ and $\dot\psi\Qpsi$, partially cancel and lead to a dip in $\PSR$. As $|\dot{\phi}|$ becomes more dominant, the cancellation fades out and $\PSR$ returns to being dictated by the newly-dominant term $\dot\phi\Qphi$.
In contrast, the entropic perturbation $S$ grows because of the opposite signs of its terms, and becomes larger than $\Rcurv$ for $-10^7 \lesssim t \lesssim -10^5$. This is the only instance in the time span explored where entropic perturbations are larger than curvature perturbations.

Progressing in time towards $t_{eq1}$, both field's influence on the background's dynamics become comparable. Both perturbations $Q^\phi$ and $Q^\psi$ react to this with a dip in their magnitude, each at their own pace.
The transition to ekpyrotic domination in the contracting phase induces a smooth \textit{phase-inversion} in the complex-valued perturbations. Before $t_{eq1}$, their imaginary part is never less than $10$ orders of magnitude larger than the real part 
(see appendix \ref{app:real-imaginary-components} for further detail regarding these non-vanishing dips). As ekpyrotic domination sets in, the imaginary component smoothly evolves from a large positive value to a large negative value, inverting its sign, while the real part remains nearly unchanged.
Consequently, the phase of each perturbation inverts from effectively $+\pi/2$ to $-\pi/2$, but the overall magnitude never actually vanishes; it reaches a non-zero minimum set by the small real component. In essence, the perturbations undergo an inversion in phase in response to the change in background domination, with the sign-reversal of the large imaginary part driving this behaviour. This phase-inversion appears as a dip in amplitude of the perturbation variables' power spectra.
This smooth phase-inversion starts off similarly for both perturbation variables shortly before $t_{eq1}$, when the ekpyrotic field becomes significant but still subleading, but is completed much faster by $\Qphi$ than by $\Qpsi$. $\Qphi$ phase-inverts from $t_{eq1}$ to $t \approx -10^{5}$, with a minimum dip in $\PSphi$ happening at $t=-6.4 \times 10^5$, while $\Qpsi$ takes from $t_{eq1}$ to right up to the bounce phase, with a minimum dip in $\PSpsi$ happening at $t=-71$.

As the bounce phase is approached, the phase-inversion behaviour of $\Qpsi$ gives way to a brief, modest amplification of its power spectrum of order $\mathcal{O}(1)$. 
After the bounce, it undergoes only minimal evolution throughout the ekpyrotic-dominated phase of expansion, again growing by only $\mathcal{O}(1)$.
Although $\Qpsi$ remains negligible for $\Rcurv$ during both ekpyrotic-dominated contraction and expansion, it still sources entropic perturbations which, nevertheless, stay orders of magnitude below the curvature perturbations.

During the ekpyrotic-dominated contraction, the rapid phase-inversion of $\Qphi$ transitions into a slow growth in amplitude, with approximately $\PSphi \propto |t|^{-3/16}$. Through the bounce, $\PSphi$ is significantly amplified by a factor of $\mathcal{O}(10)$. The larger amplification of $\Qphi$ is due to the magnitude of $\Omega_{\phi\phi}$ being far larger than the other $\Omega_{F F'}$ entering the equations of motion of the perturbation variables. This term briefly dominates the evolution of $\Qphi$ throughout the bounce phase and yields such a significant amplification.
Together with the smaller amplification of $\PSpsi$, these effects result in an overall $\mathcal{O}(10)$ increase in $\Rcurv$ through the bounce phase. After the bounce, $\Qphi$ continues to grow moderately, with approximately $\PSphi \propto |t|^{+3/16}$ during the subsequent ekpyrotic-dominated expansion, lasting until $t_{eq2}$.

As $t_{eq2}$ is approached, both fields' influence become comparable again. $Q^\psi$ is strongly amplified by $\mathcal{O}(10^2)$ as $\psi$ becomes the dominant field anew. On the other hand, the change in domination causes $Q^\phi$ to phase-invert again in a similar way as it did in the contracting phase — by its imaginary, dominant component reversing in sign. The phase-inversion is fast and only spans $t \in [\sim 10^{6}, \sim 10^{8} ]$, with a minimum dip in $\PSphi$ happening at $t=3.3 \times 10^7$. However, in this instance the phase-inversion behaviour results in the magnitude of $Q^\phi$ dropping by $\mathcal{O}(10)$. Overall, $\PSR$ only decreases by a factor of $\mathcal{O}(1)$ through the domination transition in the expanding phase. Once in the quasi-dust-dominated phase of expansion, all three $Q^\phi$, $Q^\psi$, and $\mathcal{R}$ become approximately frozen, with only a very small, negligible growth until the end of the numerical simulation. Curvature perturbations are dictated by $\Qpsi$ again, while entropic perturbations are dictated by $\Qphi$. The entropic power spectrum at $t_f$ is $6.8\%$ of the total curvature power spectrum (entropic and comoving curvature power spectra combined), such that the resulting curvature power spectrum can be said to be adiabatic.

\subsubsection{Evolution over time of the Power Spectra for \texorpdfstring{$k=10^{-2}$}{}}


\begin{figure}[t!]
    \centering
    \makebox[\linewidth]{%
        \includegraphics[width=0.57\textwidth]{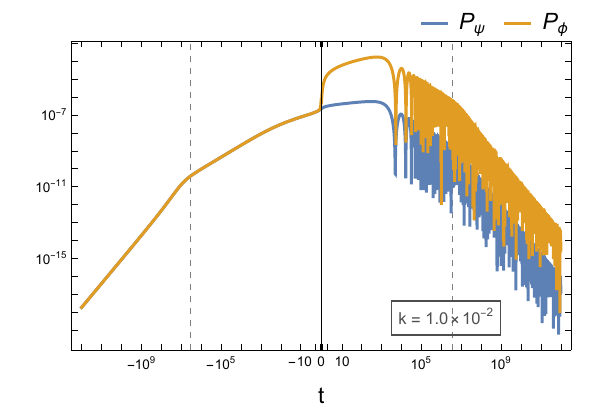}
        \hspace{-20pt}
        \includegraphics[width=0.57\textwidth]{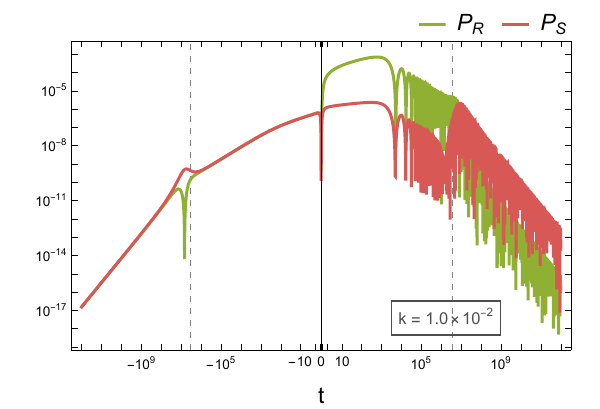}
    }
    \caption{Evolution over time of: (left) $\PSphi$ and $\PSpsi$, and (right) $\PSR$ and $\PSS$, for mode $k=10^{-2}$ in the regime that becomes oscillatory after the bounce. The vertical dashed lines mark the times of equal energy densities $t_{eq1} = -3.64 \times 10^6$ and $t_{eq2} = 3.64 \times 10^6$. The horizontal axes represent time on a signed-logarithmic scale, while the vertical axes are logarithmic. See appendix \ref{app:real-imaginary-components} for further detail regarding the oscillatory behaviour in the phase of expansion.}
    \label{fig:kSub-Scalar-tEvolution}
\end{figure}


We will now discuss the evolution of the power spectra for the mode $k=10^{-2}$.
This mode is within the horizon for most of the time span explored. It exits the horizon in the ekpyrotic-dominated phase of contraction near the bounce, it briefly reenters and re-exits the horizon throughout the bounce phase due to the vanishing of the Hubble parameter at the bounce, and finally reenters the horizon in the ekpyrotic-dominated phase of expansion briefly after the bounce.
This mode lies within the regime with a blue curvature power spectrum before the bouncing phase. The evolution of its amplitude is shown in the right plot of Figure \ref{fig:kSub-Scalar-tEvolution}.

Initially, both $\Qphi$ and $\Qpsi$ evolve with the same approximate equation of motion and from the same initial quantum vacuum state. Thus both perturbations evolve approximately identically since $t_i$, with $\PSphi \approx \PSpsi \propto |t|^{-4/3}$ throughout the quasi-dust-dominated phase of contraction. This entails that curvature and entropic perturbations are also approximately equal throughout this period for this mode, with $\PSR \approx \PSS \propto |t|^{-4/3}$.

As $t_{eq1}$ is approached, as was described for the $k=10^{-10}$ mode, $\PSR$ undergoes a dip and rebound due to partial cancellation of its terms when they are comparable in magnitude, while instead $\PSS$ grows when both fields are comparable. After the transition to ekpyrotic domination, both $\PSR$ and $\PSS$ are seen to restore the same evolution until the bounce, with all of $\PSphi$, $\PSpsi$, $\PSR$ and $\PSS$ growing at a slower rate, proportional to $|t|^{-2/3}$ in the ekpyrotic-dominated phase of contraction.

Through the bounce phase, again similar to the case of $k=10^{-10}$, $\Qpsi$ is only slightly amplified by a small factor $\mathcal{O}(1)$, while $\Qphi$ is significantly amplified by a factor $\mathcal{O}(10)$. The larger amplification of $\Qphi$, again, is due to the magnitude of $\Omega_{\phi\phi}$ being far larger than the other $\Omega_{F F'}$ entering the equations of motion of the perturbation variables.
Continuing into the ekpyrotic-dominated phase of expansion, during which $\dot\phi > \dot\psi$, curvature perturbations are dictated by $\Qphi$ and remain 3 orders of magnitude larger than entropic perturbations until $t_{eq2}$, when both fields' influence become comparable again.
Furthermore, shortly after the bounce, around $t \approx 1000$, $\Qphi$ and $\Qpsi$ become oscillatory with a decreasing averaged amplitude until the end of the time span explored 
(see appendix \ref{app:real-imaginary-components} for further detail regarding this oscillatory behaviour).
It is worth noting that after the transition to quasi-dust domination at $t_{eq2}$, the weaker perturbation variable of the pair, namely $\Qpsi$, dictates $\Rcurv$, while the stronger variable, namely $\Qphi$, dictates $\mathcal{S}$. Therefore, for $k=10^{-2}$, entropic perturbations dominate over curvature perturbations throughout the remaining expansion dominated by quasi-dust.

\subsubsection{Scalar amplitude, scalar spectral index \texorpdfstring{$n_s$}{} and its running \texorpdfstring{$\alpha_s$}{}}


\begin{figure}[t!]
    \centering
    \makebox[\linewidth]{%
        \includegraphics[width=0.52\linewidth]{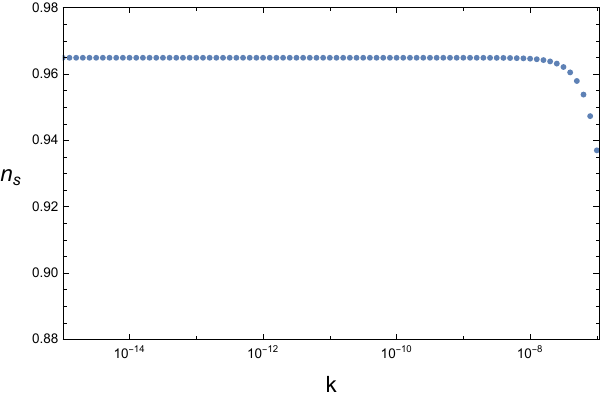}
        \hspace{6pt}
        \includegraphics[width=0.55\linewidth]{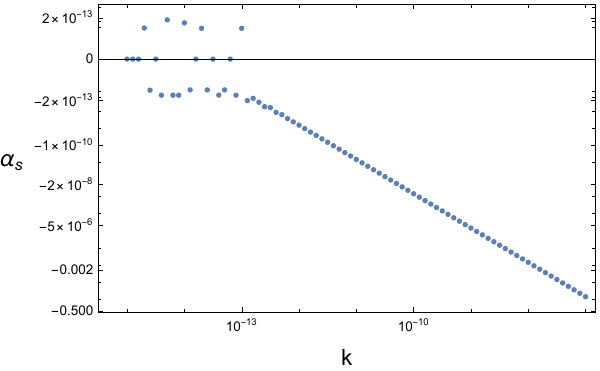}
    }
    \caption{(Left) Plot of the scalar spectral index as a function of $k$. It is approximately constant at $n_s=0.9649$ for $k<10^{-9}$, thus the power spectrum of scalar perturbations is therefore red-tilted.
    (Right) Plot of the running of the scalar spectral index as a function of $k$, with a signed-logarithmic vertical scale. The running is increasingly small in magnitude for increasingly larger modes, with $|\alpha_t| < 10^{-5}$ for $k<10^{-9}$.}
    \label{fig:Scalar-spectral-index-red-tilted}
\end{figure}


Building on the description of the evolution of the power spectra in relation to many different modes as depicted in Figure \ref{fig:power-spectra-evolution}, we now discuss the resultant scalar spectral index $n_s$ in the quasi-dust-ekpyrotic bounce scenario. We compute the scalar spectral index given in \eqref{eq:n_s-definition} using our numerical results for the curvature power spectrum $\PSR$ at $t_f$ in the expanding phase. Notably, we compute the final curvature power spectrum at $t_f$ and not $t_{eq2}$ because around the transition time the curvature power spectrum experiences a small decrease before freezing in the quasi-dust-dominated phase of expansion%
\footnote{For modes $k < 10^{-9}$ in the red-tilted regime, the curvature power spectrum decreases by $49\%$ at the onset of the quasi-dust-dominated phase of expansion before freezing at a constant value for the remaining of the time span. This can be seen for $k = 10^{-10}$ in Figure \ref{fig:kSuper-Scalar-tEvolution}.}%
. We find an approximately constant scalar spectral index at $n_s=0.9649$ for the range $k\in[10^{-15},10^{-9}]$, as can be seen in the left panel of Figure \ref{fig:Scalar-spectral-index-red-tilted}. 
Furthermore, we compute the running of the scalar spectral index as given in \eqref{eq:running-alpha_s-definition}. We find a very small, negative running with $|\alpha_s| < 10^{-5}$ for $k\in[10^{-15},10^{-9}]$, as can be seen in the right panel of Figure \ref{fig:Scalar-spectral-index-red-tilted}. The running approximately follows a power law $\alpha_s \propto k^2$ for $k\in[10^{-13},10^{-7}]$.
For $k\in[10^{-15},10^{-13}]$, the numerical precision of our computations is insufficient to accurately determine the running of the spectral index. As a result, the computed values in this range are dominated by numerical error. We anticipate that a higher-precision calculation would confirm that the running follows the same power-law behaviour observed for modes at larger $k$.
Therefore, the value found for the scalar spectral index in the red-tilted regime can be extrapolated to hold for arbitrarily small $k < 10^{-15}$.
These results are consistent with recent CMB observations which find $n_s = 0.9647 \pm 0.0044$ (68\%CL) and $\alpha_s = -0.006 \pm 0.013$ (95\%CL) when allowing for a running of the scalar spectral index and tensor perturbations \cite{Planck:2018vyg}.

Moreover, a rudimentary, back-of-the-envelope calculation enables us to further compare our results against observations.
Suppose that the cosmological scenario described in this work is followed by the radiation-dominated and matter-dominated eras of the standard hot big bang scenario until today.
The temperature of radiation is related to the scale factor by $T \sim 1/a$. If one takes the temperature at the bounce to be planckian, that is $T_\text{pl} = 1.22 \times 10^{19} \text{GeV}$, and recalls that the temperature of the CMB today is $T_\text{CMB} = 5.19 \times 10^{-13} \text{GeV}$ \cite{Fixsen:2009ug}, one finds the scale factor today to be $a_0 = 5.19 \times 10^{31}$ given that we have set the scale factor at the bounce to be $a_B = v_B^{1/3} = 1$.
The physical pivot scale for Planck measurements \cite{Planck:2018vyg} is quoted as $k^\text{P}_\text{pivot} = 0.002 \, \text{Mpc}^{-1}$, that is $k^\text{P}_\text{pivot} = 1.05 \times 10^{-60}$ in natural units. In our setting this then translates to a \textit{comoving} pivot wavenumber $k_\text{pivot} = 5.44 \times 10^{-29}$ in natural units, related via $k_\text{pivot} = a_0 \, k^\text{P}_\text{pivot}$.

We find the curvature power spectrum evaluated at $t_f$ for the pivot scale $k_\text{pivot} = 5.44 \times 10^{-29}$ to be $2.118 \times 10^{-9}$. This result is also consistent with CMB observations, which find $(2.109 \pm 0.032) \times 10^{-9}$ (68\%CL) when allowing for a running of the scalar spectral index and tensor perturbations~\cite{Planck:2018vyg}.
With regards to scalar perturbations, the scenario explored in this paper matches remarkably well with current cosmological observations.

\subsubsection{Effects of different parameter choices}

These results for the evolution of scalar perturbations are related to the choice of values for the different parameters of the model. Small variations of the ekpyrotic parameters have little impact on the evolution of perturbations. Meanwhile, varying the quasi-dust parameters does have an important effect on the resulting scalar perturbations.

As discussed in Section \ref{sec:Scalar_Initial-perturbations} there is a close relation between the value of $\epsilon$ and the resulting scalar spectral index $n_s$ given approximately by $n_s=1-12\epsilon-36\epsilon^2$ for modes that exit the Hubble horizon while $w_T \approx -\epsilon$. This relation is only approximate as evidenced by our results, but shows how the value of this parameter is strongly constrained by the requirement to match the observed red-tilted scalar power spectrum in this scenario. In this work we used this relation to find the approximate value of $\epsilon$ that would result in the observed value of $n_s$.

Moreover, as discussed in \ref{sec:Background_Effects-of-different-parameter-choices}, the value of $f$ has a strong influence in the resulting background. Different values of $f$ result in the transitions in domination between ekpyrotic and quasi-dust phases happening at different times. 
For example, decreasing the value of $f$ results in the end of the quasi-dust dominated phase of contraction, namely $t_{eq1}$, happening earlier before the bounce, at a lower energy density. The curvature perturbation grows a lot faster during the quasi-dust-dominated phase of contraction than during the ekpyrotic-dominated phase of contraction, as can be seen in Figure \ref{fig:kSuper-Scalar-tEvolution}. Thus, a longer ekpyrotic-dominated phase of contraction hinders the overall growth of the curvature perturbation in the time span explored. Consequently, decreasing the value of $f$ results in a decrease in the curvature power spectrum evaluated at $t_f$. In this work we explored the outcome of different choices of $f$ and found $f = 2.1 \times 10^{-15}$ to be a good choice to match observations of the curvature power spectrum for the pivot scale $k_\text{pivot}=5.44 \times 10^{-29}$.

\section{Dynamics of tensor perturbations and their power spectra}
\label{sec:Dynamics_Tensor_perturbations}

In this section, we discuss the tensor power spectrum from the quasi-dust-ekpyrotic bounce scenario in LQC. 
We start by laying out the Hamiltonian formalism for tensor perturbations. 
Then, we find approximate initial solutions to set in the early, classical regime, in a similar way to what we did for scalar perturbations in Section \ref{sec:scalar_perturbations}.
We similarly use the dressed metric approach to study the propagation of quantum tensor perturbations through the bounce effectively.
We then solve numerically for the evolution of the tensor perturbations, compute their power spectrum and compare it to that of scalar curvature perturbations. Finally, we discuss the results and compare them to current observations to assess the validity of the scenario.

\subsection{Tensor perturbations}

The two tensor modes that arise from the linear perturbation of the Fourier transformed spatial metric are already gauge-invariant variables since there are constraints on the tensor subspace of perturbations in the cosmological scenario being considered.
Following \cite{Pascual:2025teu}, the tensor Fourier modes for the comoving scale $k$ are collectively given by $\mathcal{T}$, where again their relation to $k$ is not expressed explicitly in the notation. The equation of motion for the tensor modes is
\be \label{eq:T-EoM}
\ddot{\mathcal{T}} + 3H \dot{\mathcal{T}} + \frac{k^2}{a^2} \mathcal{T} = 0 \;.
\ee
This equation of motion is analogous to that of two massless scalar fields, and has the same form as equations \eqref{eq:QPhi-EoM} and \eqref{eq:QPsi-EoM} after setting the different $\Omega^2_{FF'}=0$. 
Furthermore, since equation \eqref{eq:T-EoM} has the same form as the Mukhanov-Sasaki equation \eqref{eq:QF-EoM-Initially}, it follows that imposing the same initial Bunch-Davis vacuum state for tensor perturbations similarly results in the expression for the initial states set at $\eta=\eta_i=\eta(t_i)$ given by
\be \label{eq:T-solution-initially}
\mathcal{T} = \frac{1}{2a} \sqrt{-\pi (\eta - \eta_0)} \; H_\mu^{(1)} (-k(\eta - \eta_0)) \;,
\ee
and the initial speed is found to be
\be \label{eq:Tdot-solution}
\begin{split}
\dot{\mathcal{T}} = 
&-\frac{1}{2a} 
\left[
H \sqrt{-\pi (\eta - \eta_0)} 
+
\frac{1}{2a} \sqrt{-\frac{\pi}{\eta - \eta_0}}
 \; \right]
H_\mu^{(1)} (-k(\eta - \eta_0))
\\
&-\frac{k}{4a^2} \sqrt{-\pi (\eta - \eta_0)} 
\left( H_{\mu-1}^{(1)} (-k(\eta - \eta_0)) - H_{\mu+1}^{(1)} (-k(\eta - \eta_0)) \right)
\end{split}
\ee
where we recall that $\mu$ is given by \eqref{eq:Hankel-order}.

In analogy with the scalar sector, the power spectrum of tensor perturbations is given by
\be
\PST=
\frac{k^3}{\pi^2} |\mathcal{T}|^2
\ee
which accounts for both polarisations of tensor modes. In further similarity, we resort to a finite-difference approximation to the formally defined tensor spectral index for $k$ as given by
\begin{equation}
    \begin{aligned} \label{eq:n_t-definition}
        n_t(k)
        &= \frac{ \text{d} \ln\PST(k) }{ \text{d} \ln k } 
        \\
        &\approx \frac{ \ln\PST(k e^{\delta (\ln k)}) - \ln\PST(k) }{ \delta (\ln k) } \, .
    \end{aligned}
\end{equation}
and similarly for the running of the tensor spectral index $\alpha_t$, which quantifies the scale dependence of the spectral index $n_t(k)$, we use the formal definition and the finite-difference approximation given by
\begin{equation}
    \begin{aligned} \label{eq:running-alpha_t-definition}
        \alpha_t(k) 
        &= \frac{ \text{d} n_t(k) }{ \text{d} \ln k } 
        \\
        &\approx \frac{ n_t(k e^{\delta (\ln k)}) - n_t(k) }{ \delta (\ln k) } \, .
    \end{aligned}
\end{equation}
Lastly, the tensor-to-scalar ratio is defined as
\be \label{eq:tensor-to-scalar-r}
r = \frac{\PST}{\PSR} \;.
\ee

\subsection{Numerical results for tensor power spectrum}
\label{sec:Tensor_Numerical-results}


\begin{figure}[t!]
    \centering
    \hspace*{-34pt}
    \includegraphics[width=1.1\linewidth]{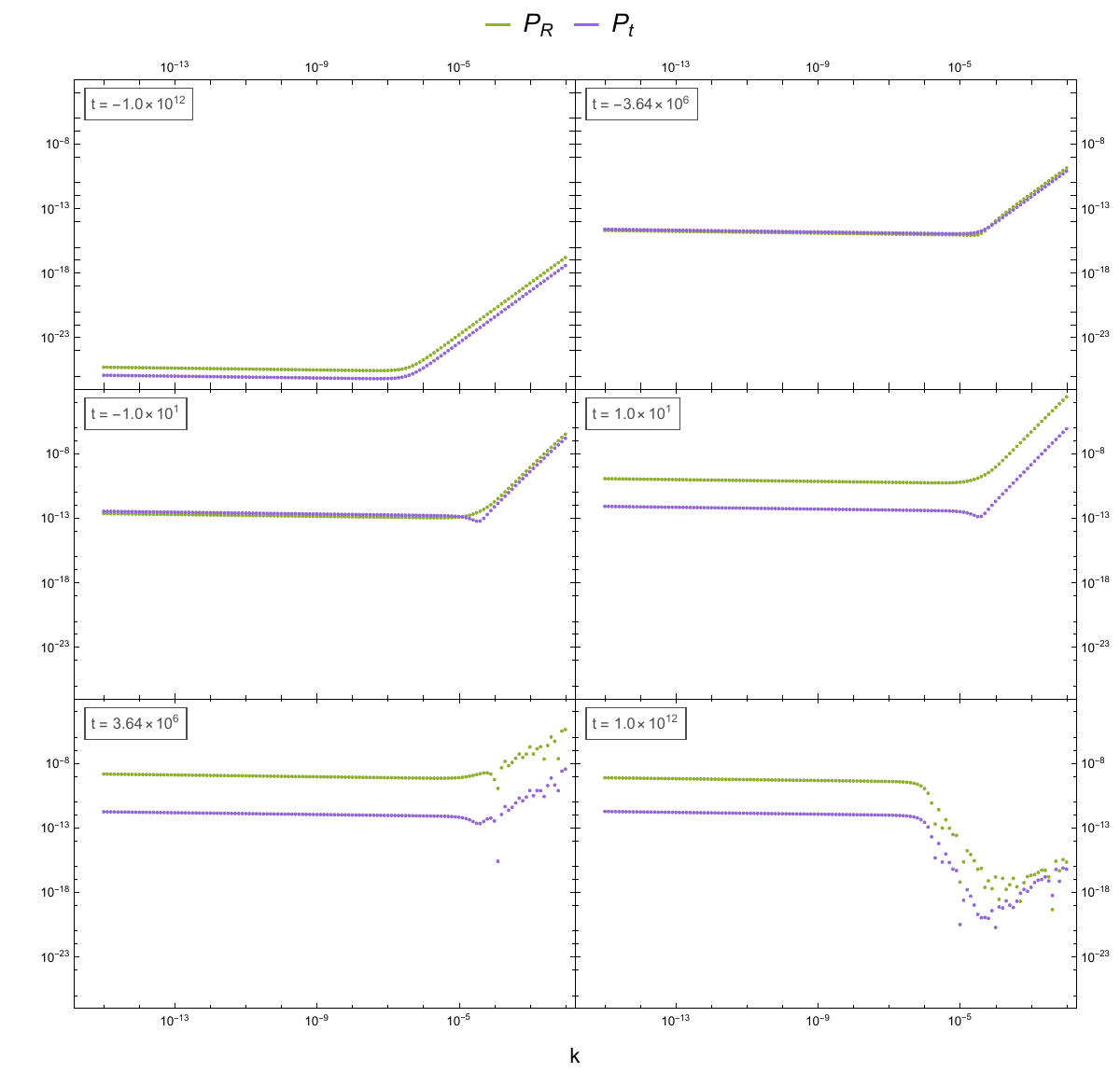}
    \caption{Evolution of the tensor ($\PST$, purple points) and curvature ($\PSR$, green points) power spectra as functions of $k$ evaluated at different times in chronological order: $t_i = -10^{12}$, $t_{eq1} = -3.64 \times 10^6$, $t = -10$, $t = 10$, $t_{eq2} = 3.64 \times 10^6$, and $t_f = 10^{12}$. Each panel shows the different power spectra evaluated for a discrete sample of uniformly logarithmically spaced comoving wavenumbers $k \in [ 10^{-15} , 10^{-2} ]$ at the labelled time. The logarithmic vertical scales are the same in all panels to facilitate comparison.}
    \label{fig:tensor-perturbations-evolution}
\end{figure}


\subsubsection{Evolution over time of the Power Spectra for different \texorpdfstring{$k$}{}}

The evolution of the amplitude of the tensor power spectrum for the range of values \linebreak $k\in[10^{-15},10^{-2}]$ is depicted in Figure \ref{fig:tensor-perturbations-evolution} along with the scalar curvature power spectrum for comparison. Initially, modes $k \lesssim 10^{-7}$ that have exited the horizon display a nearly scale-invariant power spectrum, with a slight red-tilt favouring smaller $k$. Modes $k \gtrsim 10^{-6}$ still within the horizon initially display a blue spectrum, favouring larger $k$. The tensor power spectrum has the same spectral profile as the curvature power spectrum.

Throughout the quasi-dust dominated phase of contraction, the evolution of the tensor power spectrum is very similar to the curvature power spectrum up until the transition time since the equations of motion of tensor perturbations and of the scalar perturbations are approximately the same. Their behaviour first becomes distinct around the transition to the ekpyrotic-dominated phase of contraction at $t_{eq1}$, when tensor perturbations smoothly transition to a very slow rate of growth until the bounce phase, while curvature perturbations drop slightly below tensor perturbations as a consequence of the phase-inversion undergone by $\Rcurv$ as described in section \ref{sec:Scalar_Numerical-results}.

The tensor power spectrum grows in amplitude by a very small factor $\mathcal{O}(1)$ between $t=-10$ and $t=10$ across the bounce phase, almost inappreciable in Figure \ref{fig:tensor-perturbations-evolution}, in contrast with the large amplification undergone by the curvature perturbation in this period. The bounce phase has little impact on the tensor power spectrum. As can be seen in Figure \ref{fig:tensor-perturbations-evolution}, the tensor power spectrum profile is similar to the curvature power spectrum, with red-tilted and blue regimes for the same ranges of $k$, with the exception of a dip in the tensor spectral profile for the range $10^{-5} \lesssim k \lesssim 10^{-4}$ between the red-tilted and blue regimes.

In the ekpyrotic-dominated phase of expansion after the bounce phase, the tensor power spectrum continues to increase at a very slow rate, growing only by a small factor $\mathcal{O}(1)$ until the transition to quasi-dust-dominated expansion at $t_{eq2}$. 
During this period, smaller modes in the blue regime of the power spectrum reenter the horizon in the expanding branch and acquire an oscillatory behaviour.

After $t_{eq2}$, smaller modes that reenter the horizon during the quasi-dust dominated phase of expansion acquire an oscillatory behaviour with a red amplitude. Modes that instead reentered the horizon during ekpyrotic-dominated contraction retain the blue averaged-amplitude power spectrum they acquired as they reentered. Larger modes that remain larger than the horizon in the quasi-dust dominated phase of expansion preserve their red-tilted power spectrum, and do not grow in amplitude after $t_{eq2}$.

As we did for scalar perturbations, we now focus on the time evolution for the two representative modes $k=10^{-10}$ and $k=10^{-2}$, within the red-tilted regime and the oscillatory regimes respectively. Figure \ref{fig:kSuper-kSub-Tensor-tEvolution} shows how $\PST$ evolves over time for these modes.

\subsubsection{Evolution over time of the Power Spectra for \texorpdfstring{$k=10^{-10}$}{}}


\begin{figure}[t!]
    \centering
    \makebox[\linewidth]{%
        \includegraphics[width=0.52\textwidth]{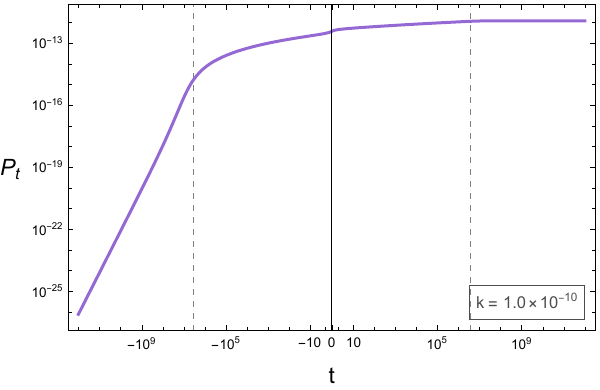}
        \hspace{6pt}
        \includegraphics[width=0.52\textwidth]{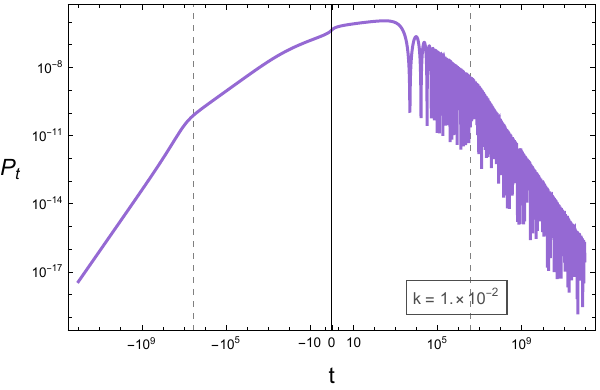}
    }
    \caption{Evolution of $\PST$ over time for different modes: (Left) mode $k=10^{-10}$ within the red-tilted regime, and (Right) mode $k=10^{-2}$ in the regime that becomes oscillatory after the bounce. The vertical dashed lines mark the times of equal energy densities $t_{eq1} = -3.64 \times 10^6$ and $t_{eq2} = 3.64 \times 10^6$. The horizontal axes represent time on a signed-logarithmic scale, while the vertical axes are logarithmic.}
    \label{fig:kSuper-kSub-Tensor-tEvolution}
\end{figure}


The mode $k=10^{-10}$ lies within the regime with a red-tilted tensor power spectrum throughout the time span explored. During the quasi-dust-dominated phase of contraction, this mode's amplitude grows fast, with $\PST \propto |t|^{-2+4w_T}$ until $t_{eq1}$. After the transition into the ekpyrotic-dominated phase of contraction, tensor perturbations continue to grow, but at a much slower rate, with approximately $\PST \propto |t|^{-3/16}$ after $t_{eq2}$ until near the bounce. As can be seen in the left panel of Figure \ref{fig:kSuper-kSub-Tensor-tEvolution}, $\PST$ undergoes a very small amplification of only $\mathcal{O}(1)$ through the bounce phase. In contrast, we recall that $\PSR$ for this mode undergoes a large amplification through the bounce phase due to the amplification of $\Qphi$. After the bounce, the rate of growth of tensor perturbations becomes very slow, with approximately $\PST \propto |t|^{1/16}$. The tensor power spectrum grows only by a factor $\mathcal{O}(1)$ during the ekpyrotic-dominated phase of expansion until $t_{eq2}$, and becomes constant as the background transitions into the quasi-dust dominated phase of expansion. Comparing Figure \ref{fig:kSuper-Scalar-tEvolution} and the left panel of Figure \ref{fig:kSuper-kSub-Tensor-tEvolution} we see that the large amplification of $\Qphi$ through the bounce is what results in $\PSR$ being larger than $\PST$ by the end of the time span explored in this scenario. 

Overall, the evolution of tensor perturbations is much simpler than that of curvature perturbations for modes in the red-tilted regime. The complexity of the scalar perturbations' evolution arises from the strong dependence of their equations of motion on the background through the terms with $\Omega_{FF'}$, including the terms coupling the evolution of the perturbations. The equation of motion for tensor perturbations is much simpler, depending only on the background through $H$ and $a$, leading in turn to a simpler evolution.

\subsubsection{Evolution over time of the tensor power spectrum for \texorpdfstring{$k=10^{-2}$}{}}

The mode $k=10^{-2}$ lies within the regime with a blue tensor power spectrum before the bounce phase. The evolution of the power spectrum for this tensor mode is shown in the right panel of Figure \ref{fig:kSuper-kSub-Tensor-tEvolution}. At this point it is worth noting that in the equation of motion for $\Qpsi$ \eqref{eq:QPsi-EoM} the terms with $\Omega_{\psi\psi}$ and $\Omega_{\phi\psi}$ are negligible compared to the term with $k^2$, as can be gathered from inspecting Figure \ref{fig:Omegas}. This implies that the equation of motion for this mode of $\Qpsi$ is in fact approximately given by \eqref{eq:QF-EoM-Initially} throughout the whole time span explored. Furthermore, recalling equation \eqref{eq:T-EoM} we gather that $\Qpsi$ and $\mathcal{T}$ for $k=10^{-2}$ undergo approximately the same evolution throughout the explored time span. Comparing the evolution of $\PSpsi$ in the left panel of Figure \ref{fig:kSub-Scalar-tEvolution} with that of $\PST$ in the right panel of Figure \ref{fig:kSuper-kSub-Tensor-tEvolution} confirms that their evolution is approximately the same
\footnote{Up to a factor of 2 in the definition of $\PST$ to account for both tensor perturbation polarisations.}.
Tensor perturbations acquire an oscillatory behaviour with a decreasing averaged amplitude when they re-enter the horizon in the expanding branch. By the end of the time span explored, the scalar curvature and tensor power spectra for this mode are found to have a similar averaged magnitude, both being 2 order of magnitude smaller than the averaged amplitude of the entropic power spectrum shown in the right panel of Figure \ref{fig:kSub-Scalar-tEvolution}.

\subsubsection{Tensor spectral index \texorpdfstring{$n_t$}{}, its running, and the tensor-to-scalar ratio \texorpdfstring{$r$}{}}


\begin{figure}[ht!]
    \centering
        \makebox[\linewidth]{%
        \includegraphics[width=0.52\linewidth]{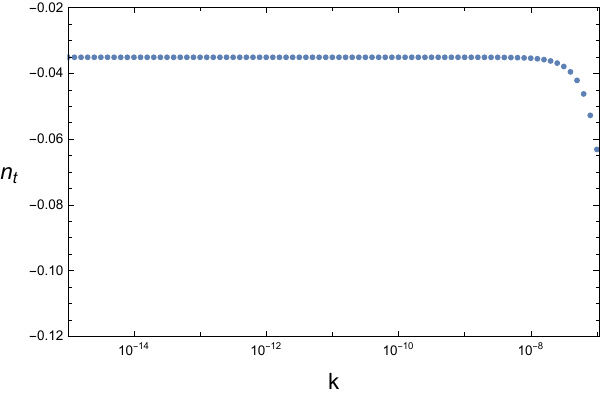}
        \hspace{6pt}
        \includegraphics[width=0.55\linewidth]{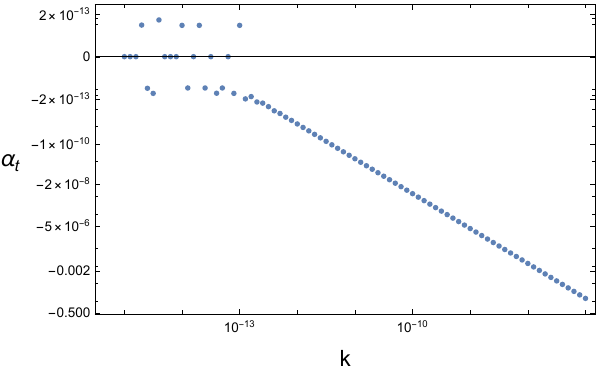}
        }
   \caption{(Left) Plot of the tensor spectral index as a function of $k$. It is approximately constant at $n_t=-0.0351$ for $k<10^{-9}$, thus the power spectrum of tensor perturbations is therefore red-tilted.
    (Right) Plot of the running of the tensor spectral index as a function of $k$, with a signed-logarithmic vertical scale. It is increasingly small in magnitude for increasingly larger modes, with $|\alpha_t| < 10^{-5}$ for $k<10^{-9}$. These show the same behaviour as their analogues for the scalar curvature power spectrum.}
    \label{fig:Tensor-spectral-index-red-tilted}
\end{figure}


We now discuss the resulting tensor spectral index $n_t$ building on the description of the evolution of the tensor power spectra in relation to many different modes as depicted in Figure \ref{fig:tensor-perturbations-evolution}.
The tensor spectral index given in \eqref{eq:n_t-definition} is computed using our numerical results for the tensor power spectrum $\PST$ at $t_f$ in the expanding phase. Nevertheless, the tensor power spectrum freezes to a constant value at $t_{eq2}$ and does not evolve anymore during the quasi-dust-dominated phase of expansion. We evaluate the tensor power spectrum at $t_f$ and not $t_{eq2}$ only to be consistent with the time of evaluation for $n_s$ carried out in Section \ref{sec:Scalar_Numerical-results}, although the tensor power spectrum is already determined at $t_{eq2}$.
We find an approximately constant tensor spectral index at $n_t=-0.0351$ for the range $k\in[10^{-15},10^{-9}]$, as can be seen in the left panel of Figure \ref{fig:Tensor-spectral-index-red-tilted}. Both tensor and curvature power spectra follow a power law with the same exponent, with the tensor and scalar spectral indices related via $n_t = n_s -1$ due to the different historical definitions of these quantities. Both tensor and curvature power spectra acquire their red-tilt for $k\in[10^{-15},10^{-9}]$ when these modes exit the horizon, even before $t_i$, in the quasi-dust dominated phase of contraction.
During this phase, the curvature perturbation is dominated by $\Qpsi$, and its equation of motion is approximately given by \eqref{eq:QF-EoM-Initially}. Notably, this matches the equation of motion \eqref{eq:T-EoM} for $\mathcal{T}$, thus yielding the same red-tilt.

Furthermore, we compute the running of the tensor spectral index as given in \eqref{eq:running-alpha_t-definition}. Similar to the scalar case, we find a very small, negative running with $|\alpha_t| < 10^{-5}$ for $k\in[10^{-15},10^{-9}]$. The running approximately follows a power law $\alpha_t \propto k^2$ for $k\in[10^{-13},10^{-7}]$.
As was for the scalar case, for $k\in[10^{-15},10^{-13}]$ the computed tensor running in this range is dominated by numerical error. We anticipate that a higher-precision calculation would confirm that the running follows the same power-law behaviour observed for modes at larger $k$.
Since inflationary models generally do not predict identical tilts for scalar and tensor modes, observing such a similarity can serve as a clear distinction between the matter-bounce and inflationary scenarios.


Lastly, we compute the tensor-to-scalar ratio evaluated at $t_f$ given in \eqref{eq:tensor-to-scalar-r} and find $r=0.00244$ for $k\in[10^{-15},10^{-7}]$. We anticipate that this ratio would hold the same value for larger scales outside the horizon $k<10^{-15}$ given the power laws found for $\alpha_s$ and $\alpha_t$ for such modes. This result is consistent with CMB observations, which find an upper bound for this ratio to be $r<0.066$ (95\%CL) when allowing for a running of the scalar spectral index and tensor perturbations for the comoving pivot scale $k_\text{pivot} = 5.44 \times 10^{-29}$ corresponding to the physical pivot scale
$k^P_\text{pivot} = 0.002 \, \text{Mpc}^{-1} = 1.05 \times 10^{-60}$ in natural units \cite{Planck:2018vyg}.

The dependence of the evolution of tensor perturbations on the choice of background parameters is similar to that of scalar perturbations. Therefore, small variations of the quasi-dust parameters cause approximately the same variation in tensor and scalar perturbations such that the effects on $n_t$ are identical to those in $n_s$.
Since both scalar and tensor red-tilted power spectra regimes vary in the same way, the resulting $r$ is mostly unaffected by variations of the model's parameters within the space that yields background solutions as specified in Section \ref{sec:Background_Dynamics}.

\section{Conclusion and Outlook}

We have introduced and analysed a two–field bouncing scenario in the setting of Loop Quantum Cosmology.
A quasi-dust scalar field with a slightly negative equation of state drives a phase of matter–dominated contraction,
while an ekpyrotic scalar field is posited to dominate and tame anisotropies early on before the bounce in order to prevent a BKL instability.
After deriving the effective background dynamics, we solved Hamilton equations of the background numerically. 
For this, we set initial conditions at the bounce and fixed a set of values for the parameters of the fields' potentials, justifying their choice, where feasible, as much as possible. 
We described the solutions for the background thoroughly, and then employed them for the evolution of the perturbations. 
After laying out the coupled equations of motion for the scalar perturbations, we set vacuum initial conditions for the perturbations deep in the quasi-dust phase of contraction and used the dressed-metric approach to solve numerically for their evolution. 
A similar procedure was later carried out for tensor perturbations, which instead do not have a coupled evolution and are simpler to deal with.

For both perturbation sectors, we performed the numerical evolution of a wide range of modes and studied how their power spectra evolved throughout the time span around the bounce. 
We then picked out two representative modes of the two regimes of modes -- those which, within the time span explored, are initially larger than the Hubble radius, and those that are initially within the Hubble radius -- and studied their evolution in detail. The evolution of modes within each respective regime is qualitatively the same, differing only in amplitude.
We presented the resulting predictions for the adiabatic curvature, entropic and tensor power spectra. 
Both scalar curvature and tensor perturbation modes that exit the Hubble radius during a phase of quasi-dust–dominated contraction acquire the same red-tilted scale-dependence and running of it.
This is due to the fact that, at the time these modes exit, their equations of motion are approximately identical, and they both start from vacuum initial conditions.
The predictions for the scalar amplitude, scalar spectral index and its running, of the curvature power spectrum generated in this model match remarkably well the current-best measurements of the primordial power spectrum inferred from the CMB. 
The parameters of the model were selected to this end. 
For the tensor power spectrum, the prediction of its amplitude is exciting: while the value of the tensor-to-scalar ratio is comfortably below the the observational upper bound, it is large enough so that it should be measured in near-future observational surveys with fast-increasing with precision. The model also predicts the scale-dependence and its running of the tensor power spectrum to be identical to that of the curvature power spectrum, which is consistent with current observational bounds, though these are not very constraining as of today. Nevertheless, the increased-precision surveys mentioned earlier may soon provide stronger constraints with which to assess the validity of this model and its predictions.

The model proposed in this work introduces a number of parameters that underpin the nature of the two scalar fields considered which, along with initial conditions for the background variables, need to be set in order to obtain numerical solutions of the background evolution. 
The parameter space for this model is large and underexplored. 
While we did not perform an exhaustive analysis of the parameter space, we investigated the outcome of a chosen set of parameters and initial background conditions -- kept as arbitrary as feasible -- to assess the suitability of the model. 
The results are encouraging, but a thorough exploration of the space of parameter values and initial conditions is paramount to assess the robustness of the scenario.

We have found that the problem of an unnatural low bouncing energy density in the matter-bounce LQC scenario (see \cite{Wilson-Ewing:2012lmx}) does not arise when there are two perturbation variables. Their coupled evolution diverts their behaviour away from their respective single-field evolution. 
Specifically curvature perturbations grow fast during matter-dominated contraction, but their growth is hindered during ekpyrotic domination. The result is that the energy scale of the bounce does not need to be reduced in order to prevent perturbations from growing much larger than the observed amplitude. Given this, it is apparent that the value of the scalar amplitude is strongly correlated to the energy scale of the transition from matter domination to ekpyrotic domination in the phase of contraction. Indeed, for the numerical study, this value (parametrised by $f$) was determined for the results to match observations.

Moreover, the amplitude of the tensor power spectrum is far lower than the curvature amplitude because, even though both grow at similar rates during ekpyrotic domination far from the bouncing phase, the latter undergoes a substantial amplification through the bouncing phase, while the former barely grows after the onset of ekpyrotic contraction. The small tensor-to-scalar ratio in this scenario is therefore an imprint of the the quantum nature of spacetime.

The next natural step in the investigation of this scenario would be to study its embedding in anisotropic spacetimes to evaluate the effectiveness of the ekpyrotic contraction regarding anisotropy suppression. It has been recently argued that, even if ekpyrotic contraction can ensure an isotropic bouncing phase, the size of anisotropies in a contracting universe is significantly more constrained by the observational bounds of the quadrupolar angular distribution induced by anisotropies in the CMB than by the scaling requirements that avoid a BKL instability \cite{Agullo:2022klq}. This needs to be addressed carefully within the matter-ekpyrotic LQC scenario. The study of the evolution of anisotropies would constrain the allowed ekpyrotic parameter values, and facilitate the study of the region of the model's parameter space that matches observations. For example, requiring that the ultra-stiff, ekpyrotic behaviour starts sooner would imply requiring a smaller value of $\alpha$. On the other hand, constraining $\alpha$ below a certain maximum value would then reduce considerably the allowed values of other parameters to have only one bounce in a given time span.

Another avenue for refinement would be to include a radiation fluid and see how it affects the evolution of perturbations. Since the amplitude of perturbations does not freeze until the ekpyrotic field becomes subdominant in the expanding phase, it would be interesting to see how differently the amplitude of super-horizon modes would evolve through a radiation-dominated contracting phase, and how the choice of parameters would vary to match observations.

While the results match observations remarkably well, the model carries conceptual shortcomings of its own. One of them is the lack of an account from LQG that can motivate a potential like the ekpyrotic given it is not readily available within established particle physics. It would mean a great conceptual feat if the mechanism that avoids a BKL instability could emerge from LQG, just like its quantisation techniques result in the resolution of singularities. One possible avenue could arise from the polymer quantisation of matter fields. Holonomisation is essentially exponentiation, which, of a matter field, could potentially lead to the sought-after ekpyrotic behaviour.

Regarding the quasi-dust field, even though its potential is conceived to capture effectively a regime where pressureless matter dominates over a tiny but non-vanishing dark energy component, the scenario would stand on a stronger ground if it did not involve such an effective description. Further investigation should explore the scenario where pressureless dust and dark energy (or a cosmological constant) are individual constituents. Another, more speculative idea is to posit that the negative pressure contribution could emerge from quantum geometry effects building up throughout the fabric of a spacetime dominated by Planckian remnants originating in the everlong, pre-bounce universe. Given that in this scenario there is no need for an inflationary phase after the bounce, Planckian remnants could have originated before the bounce and be abundant in our universe, constituting a component of dark matter.

In conclusion, we have presented a model of the early universe that can successfully account for primordial power spectra of adiabatic curvature and tensor perturbations observed. The results are positive, but further investigation into their robustness against different initial conditions and the growth of anisotropies needs to be addressed. The LQC matter-ekpyrotic scenario we have studied resolves difficulties of singular, inflationary scenarios.  Refinement, and a motivation for the scalar potentials from fundamental physics, could render it a contender for a theory of the early universe.

\section*{Acknowledgments}

The authors thank Ivan Agulló, Muxin Han, Guillermo Mena Marugán, Jérôme Quintin, Parampreet Singh, Farshid Soltani and Jared Wogan for stimulating discussions during the early phase of this work.
Research in F.V.'s research group at Western University is supported by the Canada Research Chairs Program, by the Natural Science and Engineering Council of Canada (NSERC) through the Discovery Grant ``Loop Quantum Gravity: from Computation to Phenomenology", and by the ID\# 62312 grant from the John Templeton Foundation, as part of the project \href{https://www.templeton.org/grant/the-quantum-information-structure-ofspacetime-qiss-second-phase}{``The Quantum Information Structure of Spacetime'' (QISS)}.
FV acknowledges support from the Perimeter Institute for Theoretical Physics through its affiliation program. Research at Perimeter Institute is supported by the Government of Canada through Industry Canada and by the Province of Ontario through the Ministry of Economic Development and Innovation.
The authors acknowledge the Anishinaabek, Haudenosaunee, L\=unaap\'eewak, Attawandaron, and neutral people, on whose traditional lands Western University and the Perimeter Institute are located.

\appendix

\section{Closer look at phase-inversion behaviour}
\label{app:real-imaginary-components}


\begin{figure}[th!]
    \centering
        \makebox[\textwidth]{%
            \begin{minipage}{1.15\textwidth}
                \centering
                \includegraphics[width=0.5\linewidth]{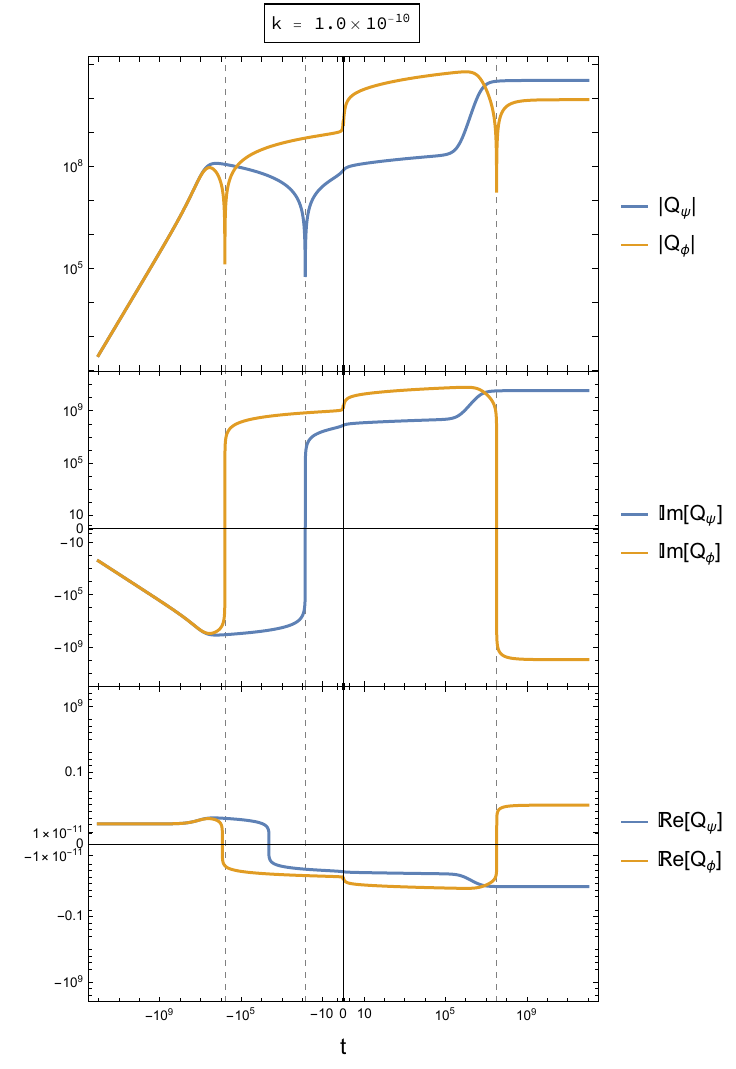}
                \hspace{-10pt}
                \includegraphics[width=0.5\linewidth]{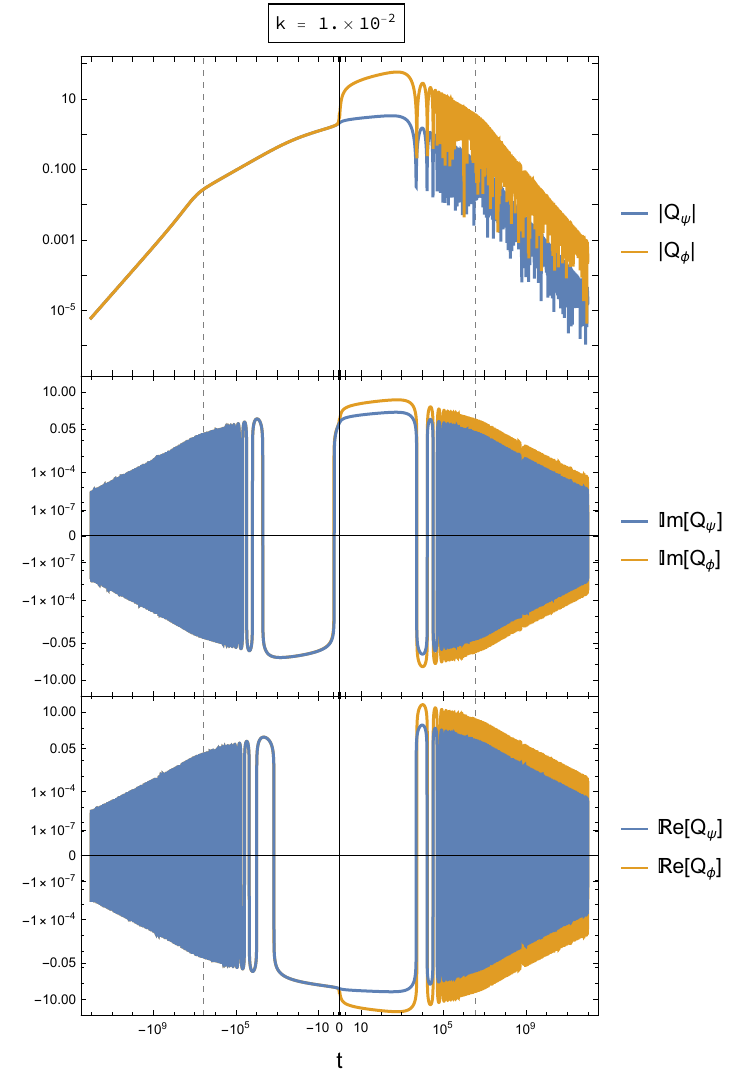}
            \end{minipage}
        }
    \caption{Evolution in time of the (top) absolute, (middle) imaginary and (bottom) real components of the scalar perturbations $\Qpsi$ and $\Qphi$ for the representative modes (left grid) $k = 10^{-10}$ in the red-tilted regime and (right grid) $k = 10^{-2}$ in the oscillatory regime.
    The dashed vertical lines in the left grid mark the times of the minima of the dips in the perturbations, at (leftmost) $t = -6.3 \times 10^5$, (middle) $t = -71$, and (rightmost) $t = 3.3 \times 10^7$. 
    The dashed lines in the right grid mark the times of equal energy densities (left) $t_{eq1} = -3.64 \times 10^6$ and (right) $t_{eq2} = 3.64 \times 10^6$.
    Note that all panels share the same signed-logarithmic time axis, but have different logarithmic scales for their vertical axes. The plotted curves often overlap. In the left grid, the blue curve is underlying the orange curve when it cannot be seen. In the right grid, the blue curve has been plotted on top of the orange in the middle and bottom panels because the orange curve has the same average amplitude in the contracting phase and a larger amplitude in the expanding phase, clearly discerned even behind the blue curve.}
    \label{fig:Abs_Im_Re_Qpsi_Qphi-grid}
\end{figure}


In this appendix we discuss in further detail the \textit{phase-inversion} behaviour of the complex-valued scalar perturbations $\Qpsi$ and $\Qphi$ by taking a closer look at the behaviours of their real and imaginary components separately.
Figure \ref{fig:Abs_Im_Re_Qpsi_Qphi-grid} shows the plots of the absolute, real and imaginary components of $\Qpsi$ and $\Qphi$ for the representative modes $k = 10^{-10}$ and $k = 10^{-2}$ discussed in section \ref{sec:Scalar_Numerical-results}.

Regarding the mode $k = 10^{-10}$ on the left side of figure \ref{fig:Abs_Im_Re_Qpsi_Qphi-grid}, in the top panel it can be seen that the absolute value of $\Qphi$ has two dips at the times $t = -6.3 \times 10^5$ and $t = 3.3 \times 10^7$, and the absolute value of $\Qpsi$ has one dip at $t = -71$. 
By means of the vertical dashed lines drawn in all three panels at the times of the dips in the absolute values, it can be seen that these dips happen at the time at which the imaginary components of the perturbations smoothly but rapidly change sign. The real components also change sign near these times, but not quite at the same time. Since the horizontal, time axes of the plots are logarithmic, differences in the times at which different components change sign become more difficult to appreciate the further away in time from the bounce they happen. 
It can be easily seen that the imaginary component of $\Qpsi$ changes sign at $t=-71$ just like the absolute value, but its real component changes sign at $t = -4.3 \times 10^3$ instead.
A close look at the (leftmost) dashed line at $t = -6.3 \times 10^5$ reveals that while for $\Qphi$ its absolute value dips and its imaginary component changes sign at this time, its real component changes sign at $t = -8.4 \times 10^5$ instead. 
Similarly, although it cannot be appreciated in the plot due to its limited resolution and the logarithmic time axis, the (rightmost) dashed line at $t = 3.3 \times 10^7$, at which the absolute value of $\Qphi$ dips and its imaginary component changes sign, does not align with the change in sign of its real component, which happens instead at  $t = 3.4 \times 10^7$.
The fact that the imaginary and real components don't cross through zero at the same time implies that the absolute value doesn't actually vanish throughout these dips. The dip in the absolute value happens because of the change of sign of the dominant imaginary component, but instead of vanishing when the imaginary component changes sign, it reaches a minimum non-zero value given by the non-vanishing real component. The real component does change sign as well, as can be seen in the left-bottom panel of figure \ref{fig:Abs_Im_Re_Qpsi_Qphi-grid}, but it has no impact on the absolute value, which is dominated by the imaginary component when this takes place.

Regarding the mode $k = 10^{-2}$ on the right side of figure \ref{fig:Abs_Im_Re_Qpsi_Qphi-grid}, it is interesting to take a closer look at the behaviour of $\Qpsi$ and $\Qphi$ components separately.
During the contracting phase, the perturbations' absolute value grows steadily until the bounce, as can be seen in the top panel. Interestingly, up until this mode exits the horizon in the ekpyrotic-dominated phase of contraction, its imaginary and real components are actually oscillating rapidly, as can be seen in the middle and bottom panels. Nevertheless, the imaginary and real components are rotating in phase, such that the absolute values do not exhibit any dips like those described for the $k = 10^{-10}$ mode.
Then, in the ekpyrotic-dominated phase of expansion, this mode re-enters the horizon, and the perturbations start rotating in phase again, however this time their absolute value also oscillates. For the middle and bottom panels of the right-hand side grid of figure \ref{fig:Abs_Im_Re_Qpsi_Qphi-grid} we have used the same vertical scales so that it can be appreciated that during the contracting phase, the real and imaginary components of the perturbations share a similar amplitude and neither can be said to be the dominant contribution to the absolute value. Having the same amplitude is what leads to the absolute value being non-oscillatory and following piece-wise power laws at different stages. However, through the bounce the real components are amplified further with respect to the imaginary components, such that they come to dominate the absolute values of the perturbations. The absolute value then inherits the oscillatory behaviour from the dominant real component, which is no longer matched by the imaginary component. This contrasts the case for $k = 10^{-10}$, for which the imaginary component is always dominant.

As discussed in section \ref{sec:Tensor_Numerical-results}, for the mode $k = 10^{-2}$ the evolution of tensor perturbations is qualitatively the same as that of $\Qpsi$. Therefore, its real and imaginary components share a similar behaviour as well. We do not include their respective plots because they do not add further insight.
Furthermore, the behaviour of the scalar perturbations $\Rcurv$ and $\mathcal{S}$ is dictated by that of $\Qpsi$ and $\Qphi$, including that of their imaginary and real components. A closer look at their components does not reveal any further insight, thus we do not include their plots in this discussion.

\section{Evolution of perturbations for \texorpdfstring{$k_\text{pivot}$}{}}
\label{app:kPivot}


\begin{figure}[ht!]
    \centering
    \makebox[\textwidth][c]{%
        \begin{minipage}{1.15\textwidth}
            \centering
            \includegraphics[width=0.51\linewidth]{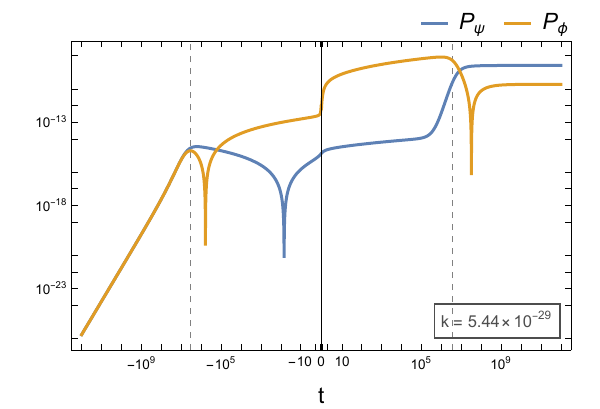}
            \hspace{-20pt}%
            \includegraphics[width=0.51\linewidth]{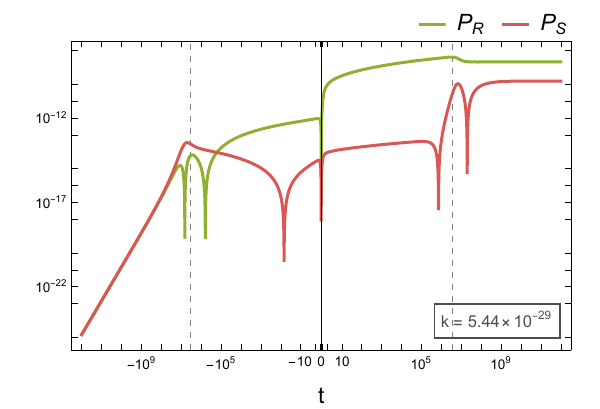}
            \\[12pt]
            \includegraphics[width=0.49\linewidth]{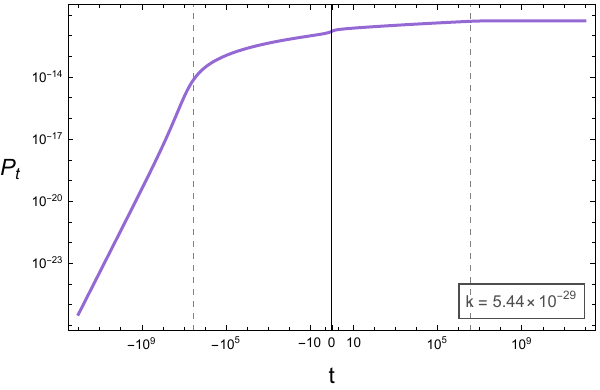}
        \end{minipage}
    }
    \caption{Evolution in time of: (top-left) $\PSphi$ and $\PSpsi$, (top-right) $\PSR$ and $\PSS$, and (bottom) $\PST$ for the pivot scale $k_\text{pivot} = 5.44 \times 10^{-29}$. This mode is within the red-tilted regime and has the same qualitative evolution as other modes in this regime. The horizontal axes represent time on a signed-logarithmic scale, while the vertical axes are logarithmic. The vertical dashed lines mark the times of equal energy densities $t_{eq1} = -3.64 \times 10^6$ and $t_{eq2} = 3.64 \times 10^6$.}
    \label{fig:kPivot-Scalar-and-Tensor-tEvolution}
\end{figure}


In this appendix we include the plots of the evolution of the scalar and tensor perturbations for the pivot scale $k_\text{pivot} = 5.44 \times 10^{-29}$ as discussed in section \ref{sec:Scalar_Numerical-results} and \ref{sec:Tensor_Numerical-results}, contained in figure \ref{fig:kPivot-Scalar-and-Tensor-tEvolution}.
It can be seen that the plots are qualitatively the same as for $k=10^{-10}$. As was discussed in section \ref{sec:Scalar_Numerical-results}, all modes in the red-tilted regime have the same qualitative behaviour. We include these plots here simply as further reference since this is the mode that was used to match its scalar power spectrum's amplitude and tilt with observations in order to select the values of the different parameters in the model. We do not include these plots in section \ref{sec:Scalar_Numerical-results} because this mode is far smaller than the sample range of modes considered there. Enlarging the range of modes explored in section \ref{sec:Scalar_Numerical-results} to include this value was deemed to computationally expensive and would not provide any further insight. The relevant insight about the scalar and tensor spectral indices and their runnings can be concluded with the range explored, namely $k \in [ 10^{-15} , 10^{-2} ]$.

\newpage
\bibliographystyle{utphys-modified}
\bibliography{mateo_libraryBBT}

\end{document}